\documentclass[10pt,prx,aps,amssymb,twocolumn,superscriptaddress, notitlepage,longbibliography]{revtex4-2}

\usepackage{mathtools}
\usepackage{graphicx} 
\usepackage{xcolor}
\usepackage[colorlinks = true,
            linkcolor = blue,
            urlcolor  = blue,
            citecolor = blue,
            anchorcolor = blue]{hyperref}
\usepackage[normalem]{ulem}
\usepackage{physics}
\usepackage{amsthm}
\usepackage{orcidlink}
\usepackage{appendix}
\usepackage{amsfonts}

\usepackage{pgfplots}
\usepackage{amsmath}
\usepackage{comment}
\usepackage{ifthen}
\usepackage{enumitem}

\usepackage{tikz}
\usepackage{quantikz}
\usetikzlibrary{calc}
 
\definecolor{tensorblue}{RGB}{100, 149, 210}
\definecolor{tensorred} {RGB}{196,  80,  78}
 
\pgfmathsetmacro{\sc}{0.5}
 
\pgfmathsetmacro{\Step}   {\sc * 1.30}   
\pgfmathsetmacro{\LegLen} {0.72 * \Step} 
\pgfmathsetmacro{\Boxsz}  {\sc * 0.83}   
\pgfmathsetmacro{\Bw}     {\sc * 2.20}   
\pgfmathsetmacro{\Lw}     {\sc * 1.80}   
\pgfmathsetmacro{\Nw}     {\sc * 1.00}   
\pgfmathsetmacro{\Rc}     {\sc * 4.00}   
\pgfmathsetmacro{\Nx}     {\sc * 0.17}   
\pgfmathsetmacro{\Ny}     {\sc * 0.27}   
\pgfmathsetmacro{\Na}     {\sc * 0.13}   
 
\tikzset{
  tbox/.style = {draw, line width=\Bw pt, rounded corners=\Rc pt,
                 minimum size=\Boxsz cm, fill=#1},
  tbox/.default = tensorblue,
  leg/.style    = {line width=\Lw pt, line cap=round},
  lbl/.style    = {font=\large\itshape},
}
 
\usepackage{tikz-3dplot}
\usetikzlibrary{calc}
\usetikzlibrary{intersections,shapes.arrows}
\pgfplotsset{compat=1.17}

\tdplotsetmaincoords{60}{30}

\newcommand{\rev}[1]{\textcolor{blue}{#1}}
\newcommand{\unilu}{Department of Physics and Materials Science, University of Luxembourg, L-1511 Luxembourg, Luxembourg}

\begin{document}      

\title{Shortcuts to Adiabaticity for non-Hermitian systems in Krylov Space}


\author{Ankit W. Shrestha \orcidlink{0009-0008-5367-7062}}
    \email{ankit.wenjushrestha@uni.lu} 
    \affiliation{\unilu}

\author{Budhaditya Bhattacharjee \orcidlink{0000-0003-1982-1346}}
\email{budhaditya.bhattacharjee@uni.lu}
    \affiliation{\unilu}

\author{Adolfo~del Campo\orcidlink{0000-0003-2219-2851}} 
    \email{adolfo.delcampo@uni.lu}
    \affiliation{\unilu}
    \affiliation{Donostia International Physics Center, E-20018 San Sebastián, Spain}

\date{\today}

\begin{abstract}
Shortcuts to adiabaticity (STA) reproduce adiabatic dynamics in finite time, but their counterdiabatic implementation relies on the adiabatic gauge potential (AGP), which is difficult to compute and implement in many-body systems and whose extension to open and non-Hermitian settings has remained largely model-specific. Here, we develop a general, diagonalization-free framework for engineering STA in non-Hermitian systems by representing the AGP in Krylov space. Starting from an integral representation of the counterdiabatic control, we recast the AGP as a nested-commutator series with controlled locality and generate the associated Krylov basis using the bi-Lanczos and Arnoldi algorithms. This reduces the exact or truncated AGP to a sparse tridiagonal or upper-Hessenberg matrix equation that generalizes the Hermitian construction. We demonstrate the method on a decaying two-level atom, where it recovers the exact drive and signals the exceptional point; on the interacting Hatano-Nelson model, where truncated controls rapidly suppress nonadiabatic excitations; and on a $\mathcal{PT}-$symmetric Heisenberg chain, whose AGP norm detects the $\mathcal{PT}-$symmetry-breaking transition. Throughout, the expansion converges with only a small fraction of the full Krylov space, offering a practical route to fast, accurate control of many-body non-Hermitian systems. 
\end{abstract}

\maketitle

\section{Introduction}
Shortcuts to Adiabaticity (STA) make it possible to accelerate a process that follows an adiabatic reference trajectory in a nonadiabatic fashion \cite{Chen2010}. To that end, STA generally requires auxiliary counterdiabatic control fields that modify the system Hamiltonian \cite{Demirplak&Rice_2003,Demirplak&Rice_2005,Demirplak&Rice_2008,Berry_2009}. Initially thought to be confined to single-particle systems, STA have now been theoretically developed and experimentally demonstrated across a wide variety of scenarios \cite{Torrontegui2013,GueryOdelin2019,Hatomura2024}. In particular, the experimental implementation of STA has been demonstrated both at the single- and many-particle level \cite{Visuri2026}.

Much of this progress has been fostered by the realization that CD controls are generally difficult to implement, particularly in many-body systems, where they involve nonlocal many-body interactions \cite{delCampo2012,Takahashi2013,Damski2014}. This motivated the development of approximate controls using variational methods \cite{Takahashi2013,Saberi2014,Takahashi2015,Sels2017}. In isolated systems, an integral representation of the CD term \cite{Claeys_2019} enabled nested-commutator expansions with controlled locality and the use of Krylov subspace methods \cite{Claeys_2019,Takahashi_2024STA,Bhattacharjee_2023,Morawetz2025,Grabarits2025}. These approaches have since been adopted in quantum algorithms for optimization and state preparation \cite{Chandarana2022,Hegade2022} and implemented on several quantum platforms \cite{Hegade2021,Chandarana2023,Chandarana2024,Visuri2026}.

By contrast to isolated systems, progress in controlling open quantum systems is far more limited. System-specific approaches have been put forward \cite{Dann2019,Dupays2020,Dupays2021}, and an experimental demonstration has been reported in superconducting circuit quantum electrodynamics \cite{Yin2022}. General schemes, however, require tailoring the coupling to the environment \cite{Vacanti2014,Alipour2020shortcutsto,Alipour2022}, and their experimental implementation remains elusive. The use of measurements has been proposed as an alternative \cite{Tanaka2012,HacohenGourgy18,Lewalle2024}, leveraging the adaptive quantum Zeno effect \cite{delcampo2026}.

Among the class of open quantum systems, non-Hermitian systems offer significant advantages. Theoretically, their description in terms of non-Hermitian Hamiltonians allows for analogous approaches to those developed in the Hermitian setting \cite{Ashida_NHPhysics}. Experimentally, non-Hermitian Hamiltonian systems can generally be engineered by postselection of continuously monitored quantum systems in the no-click limit. However, they also arise as an effective description in the optical platforms.  STA for open quantum systems were introduced in \cite{Ibanez2011,Ibanez2012Erratum} and subsequent developments \cite{Torosov2013,Hornedal2025}. Recently, CD protocols have been experimentally demonstrated in an effective non-Hermitian qubit using continuous monitoring in a superconducting system \cite{Erdamar2026}.  At the many-particle level, non-Hermitian systems exhibit very rich physics \cite{Ashida_NHPhysics} and their control is desirable for applications in quantum science and technology, including optimization \cite{McArdle2019,Motta2020} and sensing \cite{Lau2018,McDonald2020,Ding2023}. 
Shortcuts in many-particle non-Hermitian systems have only been found theoretically in model-specific scenarios \cite{Dupays2025}.
Yet, harnessing many-body effects by counterdiabatic controls remains crucial for many applications.

In this work, we introduce a framework for engineering shortcuts to adiabaticity in non-Hermitian systems. We first derive an integral representation of the counterdiabatic controls, which we then use to construct a nested-commutator expansion with controlled locality. We demonstrate the method on a variety of systems, including a decaying two-level atom, the interacting Hatano-Nelson model, a $\mathcal{PT}-$symmetric Heisenberg chain and the $\mathcal{PT}-$symmetric Non-Hermitian Transverse field Ising model. We demonstrate the efficacy of the Krylov approach towards the detection of quantum phase transitions in $\mathcal{PT}-$symmetric quantum systems.
In doing so, we establish a new avenue for controlling non-Hermitian many-body systems in Krylov space.

\section{Preliminaries}

We start by briefly reviewing the engineering of STA by counterdiabatic driving in Hermitian systems \cite{Demirplak&Rice_2003,Demirplak&Rice_2005,Demirplak&Rice_2008,Berry_2009}. This approach utilizes the geometric nature of control Hamiltonian to minimize transitions among different energy levels. Consider a Hamiltonian $\mathcal{H}[\lambda(t)]$ where $\lambda$ is a tunable parameter that can be varied with time. We will denote $\lambda(t) \equiv \lambda$ for brevity. The instantaneous eigenbasis of the Hamiltonian is given by
\begin{align}
    \mathcal{H}(\lambda) \ket{n(\lambda)} = \mathcal{E}_n(\lambda) \ket{n(\lambda)}\,, \label{eig_val_eqn}
\end{align}
where the Hermiticity of $\mathcal{H}(\lambda)$ ensures $\mathcal{E}_n \in \mathbb{R}$ and the basis states $\{\ket{n(\lambda)}\}$ are complete and orthogonal. An arbitrary state $\ket{\psi(\lambda)}$ satisfies the time-dependent Schr\"{o}dinger equation (setting $\hbar = 1$)
\begin{align}
    i\dot{\lambda} \frac{d}{d \lambda}\ket{\psi(\lambda)} = \mathcal{H}(\lambda)\,\ket{\psi(\lambda)}\,.
\end{align}
If we assume adiabatic driving of the eigenstates such that they only accumulate a phase upon time evolution, the state at a later time is given by $\ket{\psi_{\rm ad}(\lambda)} = \sum_n c_n e^{i\gamma_n}\ket{n(\lambda)}$. The phase $\gamma_n$ can be obtained from  
\begin{align}
    \gamma_n = -\int_{0}^{t} \mathcal{E}_n(t')\,\mathrm{d}t' + i \int_{0}^{t}\braket{n}{\partial_{t'} n} \,\mathrm{d}t'\,,
\end{align}
where the two contributions come from the dynamical phase and the Berry phase, respectively. The constraint for slow variation of the parameter to ensure adiabatic (or transitionless) driving can be circumvented by introducing a counterdiabatic (CD) term to the Hamiltonian, such that it satisfies
\begin{align}
        i\dot{\lambda} \frac{d}{d \lambda}\ket{\psi_{\rm ad}(\lambda)} = (\mathcal{H}(\lambda) + \mathcal{H}_{\rm CD}(\lambda))\,\ket{\psi_{\rm ad}(\lambda)}\,. \label{CD_eqn}
\end{align}
Plugging in $\ket{\psi_{\rm ad}(\lambda)}$ to Eq. \eqref{CD_eqn}, we get the explicit form of the CD Hamiltonian 
\begin{align}
    \mathcal{H}_{\rm CD} &= \dot{\lambda}i
    \sum_n \Big[\ket{\partial_\lambda n} \bra{n} - \braket{n}{\partial_\lambda n} \ket{n} \bra{n}\Big] \nonumber\\
    &\equiv \dot{\lambda} A_\lambda\,,
\end{align}
where $\ket{n}=\ket{n(\lambda)}$ and  $A_\lambda$ is the Adiabatic Gauge Potential (AGP). The above expression is not unique as there is ambiguity in the phase factor in $\ket{n(\lambda)}$. If we instead chose $\ket{\tilde{n}} = e^{-\int \braket{n}{\partial_\lambda n} d\lambda} \ket{n}$ to absorb the Berry phase, the AGP in the new basis becomes
\begin{align}
    \tilde{A}_\lambda = i \sum_n\ket{\partial_\lambda\tilde{n}}\bra{\tilde{n}} = i\partial_\lambda \label{AGP}\,.
\end{align}

The expression in \eqref{AGP} can be derived alternatively from the formulation of Demirplak and Rice in \cite{Demirplak&Rice_2003, Demirplak&Rice_2005, Demirplak&Rice_2008}, which utilizes the transformation into the instantaneous eigenbasis of $H(\lambda)$ generated by a unitary $U_\lambda$. The additional terms are then added to cancel any off-diagonal contribution in this basis (see \cite{pandey2021studies} for a pedagogical discussion). 

The matrix elements of the AGP can be written using the Feynman-Hellmann theorem \cite{Feynman_1939} as
\begin{align}
    \bra{m}A_\lambda\ket{n} = -i \frac{\bra{m}\partial_\lambda \mathcal{H} \ket{n}}{\mathcal{E}_m - \mathcal{E}_n}\;\; \forall\, m \neq n\,. \label{AGP_matrix_hermitian}
\end{align}
The AGP also satisfies the commutation relation \cite{Kolodrubetz_2017}
\begin{align}
    [\mathcal{H}, i\partial_\lambda \mathcal{H}+[\mathcal{H}, A_\lambda]] = 0\,. \label{AGP_comm}
\end{align}
or equivalently, $\mathcal{\hat{L}_\lambda}[i\partial_\lambda \mathcal{H}(\lambda) + \mathcal{\hat{L}}_\lambda A(\lambda)] = 0$ with $\mathcal{\hat{L}_\lambda}(\cdot) = [\mathcal{H}_\lambda, \cdot]$ \cite{Takahashi_2024STA}. The phase ambiguity in this form is reflected as the gauge freedom in AGP, where its matrix representation $A_\lambda \to A_\lambda + K$ as long as $[\mathcal{H},K] = 0$. The matrix representation in \eqref{AGP_matrix_hermitian} highlights the problems in calculating AGP for many-body systems \cite{delCampo2012,Takahashi2013,Saberi2014,Damski2014}: it requires exact diagonalization to get instantaneous eigenstates and any small energy gap ($\mathcal{E}_m - \mathcal{E}_n$) may result in divergent terms. An alternative form of AGP was proposed in \cite{Claeys_2019} involving nested commutators
\begin{align}
    A_\lambda &=i \sum_k \alpha_k(\lambda)\underbrace{[\mathcal{H}, [\mathcal{H},\dots[\mathcal{H}}_{2k-1}, \partial_\lambda \mathcal{H}]]] \nonumber\\
    &\equiv i\sum_k \alpha_k(\lambda) \mathcal{L}^{2k-1}\partial_\lambda \mathcal{H}\,,
\end{align}
where the finite truncation results in an approximate AGP. Krylov basis, generated from the iterative Lanczos algorithm, provides an efficient mechanism for truncating the infinite sum \cite{Takahashi_2024STA, Bhattacharjee_2023}.

\subsection{STA for Non-Hermitian Systems}
The theory of STA by CD was extended to weak non-Hermitian systems in \cite{Ibanez2011, Ibanez2012Erratum} using the biorthogonal basis \cite{Muga2004,Ashida_NHPhysics}. For a general non-Hermitian Hamiltonian $H(\lambda)$, we can use a set of right $\{\ket{R_n(\lambda)}\}$ and left $\{\ket{L_n(\lambda)}\}$ eigenvectors such that they satisfy
\begin{align}
   H(\lambda) \ket{R_n(\lambda)} = E_n(\lambda) \ket{R_n(\lambda)}\,, \nonumber\\
    H^\dagger(\lambda) \ket{L_n(\lambda)} = E_n^*(\lambda) \ket{L_n(\lambda)}\,,\label{time-indep schrodinger eqn}
\end{align}
and the orthonormality condition 
\begin{align}
    \braket{{L_m}(\lambda)}{R_n(\lambda)} = \delta_{mn}\,.
\end{align}
An arbitrary state $\ket{\psi_R}$ and its biorthogonal complement $\ket{\psi_L}$ (such that $\braket{\psi_L}{\psi_R} = 1$) satisfy the time-dependent Schr\"{o}dinger equations 
\begin{align}
    i \dot{\lambda}\partial_\lambda \ket{\psi_R(\lambda)} &= H(\lambda)\ket{\psi_R(\lambda)}\,, \nonumber \\
    i \dot{\lambda}\partial_\lambda \ket{\psi_L(\lambda)} &= H^\dagger(\lambda)\ket{\psi_L(\lambda)}\,. \label{time-dep schrodinger eqn}
\end{align}
Such Hamiltonians can be diagonalized as long as their geometric multiplicity is equal to their algebraic multiplicity. There may exist some points in the parameter space $\Lambda \equiv \{\lambda\}$, called exceptional points (EP), where the condition is not satisfied and the Hamiltonian cannot be diagonalized \cite{Zhang_2019}.

The energy values $\{E_n\}$ are not real in general, and this imposes a constraint on the adiabatic theorem. Focusing purely on the dynamical phase, the adiabatic state is 
\begin{align}
    \ket{\psi^{\rm ad}_R(t)} = \sum_n c_n\, {\rm{exp}} \bigg[{-i\int_{0}^{t}E_n(t')dt'}\bigg]\,\ket{R_n}\,.
\end{align}
Plugging the above expression into \eqref{time-dep schrodinger eqn} and using the eigenvalue relations \eqref{time-indep schrodinger eqn} gives the condition
\begin{align}
    \sum_n \dot{\lambda}\Big(\partial_\lambda c_n \ket{R_n} + c_n \ket{\partial_\lambda R_n}\Big)\, {\rm{exp}} \bigg[{-i\int_{0}^{t}E_n(t')dt'}\bigg] = 0\,.
\end{align}
We can further multiply with the left eigenvector $\bra{L_m}$, and use the orthonormality condition to get
\begin{align}
    &\partial_\lambda c_m = -c_m\braket{L_m}{\partial_\lambda R_m} \\
    &- \sum_{n \neq m} c_n \braket{L_m}{\partial_\lambda R_n} {\rm{exp}} \bigg[-i\int_{0}^{t}\left[E_n(t') - E_m(t')\right]d t'\bigg] \,.\nonumber  
\end{align}
The second term can grow exponentially for the case of complex $\{E_j\}$, and thus cannot be neglected \cite{Zhang_2019}. The usual adiabatic condition 
\begin{align}
    \dot{\lambda}\left|\frac{\braket{L_m}{\partial_\lambda R_n}}{E_m - E_n} \right| \ll 1\;\; \forall\, m\neq n
\end{align}
is no longer sufficient for transitionless driving if the energy eigenvalues are complex. The validity of the adiabatic theorem for non-Hermitian Hamiltonians with a real spectrum is proved rigorously in \cite{Huang2025adiabatic}. For weak non-Hermitian systems satisfying
\begin{align}
   {\rm{exp}} \bigg[{\int_{0}^{t}{\rm Im}(E_n(t') - E_m(t'))dt'}\bigg] \approx 1\,, \label{weak non-Hermiticity}
\end{align}
it was shown in \cite{Ibanez2011} that the counterdiabatic driving term can be written using the biorthogonal basis as
\begin{align}
    H_{\rm CD} &= 
    \dot{\lambda}i
    \sum_n \Big[ \ket{ \partial_\lambda R_n}\bra{L_n}  - \braket{L_n}{\partial_\lambda R_n} \ket{R_n}\bra{L_n}\Big]\, \nonumber \\ &\equiv \dot{\lambda}A_\lambda\, .
\end{align}
The final term can be dropped with an appropriate choice of gauge. It is equivalent to choosing a basis where diagonal entries are zero. In the remainder of the manuscript, we denote the tuning parameter by $\lambda$, which is implicitly dependent on time $t$. 

\section{Integral Representation of the AGP}

The Adiabatic Gauge Potential in the biorthogonal basis can be written equivalently as
\begin{align}
    A_\lambda = -i\sum_n \ket{R_n} \bra{\partial_\lambda L_n}\,.
\end{align}
When the Hamiltonian is nondegenerate, it admits the explicit matrix form
\begin{align}
    \bra{L_m}A_\lambda \ket{R_n} = -i\frac{\bra{L_m}\partial_\lambda H \ket{R_n}}{E_m - E_n}\;\; \forall\,m \neq n\,. \label{Matrix_AGP}
\end{align}
The numerator on the right-hand side can be understood in terms of the Feynman-Hellmann theorem generalized to non-Hermitian systems \cite{Hajong2024HFtheorem}. The above form of the AGP admits an integral representation over fictitious time $s$ (see Appendix \ref{Integral_rep})
\begin{align}
    A_\lambda = -\frac{1}{2} \lim_{\mu \to 0} \int_{-\infty}^\infty {\rm sgn}(s)e^{-\mu |s|}\, e^{iH_\lambda s} \partial_\lambda He^{-iH_\lambda s}\, ds\,. \label{integral_AGP}
\end{align}
This representation is structurally different from the integral representation for Hermitian systems~\cite{Sels2017}. The physically relevant time evolution in non-Hermitian systems is generated by $U^\dagger A U$ for any operator $A$ and time-evolution matrix $U$ (expressed as $U = \exp(-i t H)$, which makes it non-unitary). For a density matrix $\rho$, this implies the following Liouville-von Neumann equation
\begin{align}
    \frac{\mathrm{d}\rho}{\mathrm{d}t} = -i(H\rho-\rho H^\dagger)\,,
\end{align}
as a consequence of the evolution $\rho(t) = U \rho U^\dagger$. In the integral representation in Eq.~\eqref{integral_AGP}, the dynamics is instead generated by $U^{-1} A U$, and the von Neumann and Heisenberg equations are identical to the Hermitian version. The integral representation in Eq.~\eqref{integral_AGP} involves \rev{an} \emph{isospectral flow}, in contrast to the Heisenberg evolution of a non-Hermitian system. 

The integral representation can also be cast in terms of the following constraint equation
\begin{align}
\mathcal{L_\lambda}[i\partial_\lambda H(\lambda) + \mathcal{L}_\lambda A(\lambda)] = 0\,, \label{AGP_commutator_repn}
\end{align}
with $\mathcal{L_\lambda}(\cdot) = [H_\lambda, \cdot]$. Using the Baker-Campbell-Hausdorff expansion, the AGP can be written in terms of a series of nested commutators 
\begin{align}
A_\lambda = i\sum_k \tilde{\alpha}_k (\lambda) \mathcal{L}_\lambda^{2k-1}\partial_\lambda H_\lambda\,, \label{AGP_nested}
\end{align}
if it has a spectral gap \cite{Claeys_2019}. As discussed earlier, although the most natural basis to write AGP is in the eigenbasis of the Hamiltonian, it requires exact diagonalization, which may not always be feasible. Another efficient alternative, motivated by the form of Eq. \eqref{AGP_nested}, is to use the Krylov basis generated by repeated action of $\mathcal{L}$ on $\partial_\lambda H$.

\section{Krylov Expansion} \label{AGP_Krylov_Section}

The form of AGP in Eq. \eqref{AGP_nested} clearly indicates that it only involves nested commutators of $H$ with $\partial_\lambda H$. The inverse dependence on frequency $\omega_{m n} = E_{m} - E_{n}$ implies that the Fourier expansion of the AGP can only contain odd powers of $\omega_{m n}$, in turn enforcing that even nested commutators cannot contribute to the AGP. This is represented in Eq.~\eqref{AGP_nested}. For a finite dimensional system, this infinite nested commutator series can be repackaged into a finite series by choosing an appropriate orthonormal expansion~\cite{Hatomura2021Controlling}. Thus, the Krylov basis generated from the iterative action of $\mathcal{L}_\lambda$ on $\partial_\lambda H$ provides an efficient set of orthonormal vectors to represent $A_\lambda$. 

The general idea of the Krylov expansion involves developing an orthonormal basis $P_k$ (or equivalently, $\vert P_{k} )$ in vectorized notation, which we adopt later). The basis is defined by the initial operator $\partial_\lambda H$, its time evolution generated by $\mathcal{L}_\lambda$ and an inner product, e.g., the Hilbert-Schmidt inner product $(A\vert B) = \Tr(A^\dagger B)$. The method involves recursively generating the basis by orthonormalizing the nested commutators. Interested readers can refer to \cite{Pratik2025Krylovreview, rabinovici2025krylovcomplexity} for a more general overview. 

To evaluate the dynamics under $U^{-1}\partial_\lambda H U$, we employ the  \emph{bi-Lanczos} and \emph{Arnoldi} algorithms. The bi-Lanczos  algorithm is discussed in detail in Appendix~\ref{sec:bilanczos}, while the  Arnoldi algorithm is discussed in Appendix~\ref{sec:arnoldi}. The bi-Lanczos algorithm generates a bi-orthogonal basis $|Q_k), |P_k)$ such that the generator $\mathcal{L}_\lambda$ becomes tridiagonal in this basis
\begin{align}
    (Q_m \vert \mathcal{L}_\lambda \vert P_n) = a_{m}\delta_{m,n} + b_{n}\delta_{m,n-1} + c_{n+1}\delta_{m,n+1}\,.\label{eq:bilanczos-alg-1}
\end{align}
Moreover, we only need to consider the Krylov basis at odd order $\{|P_{2k-1})\}$ given that only $\mathcal{L}^{2k-1}|\partial_\lambda H)$ appears in the series expansion. We can write $|A_\lambda)$ in the Krylov basis as
\begin{align}
    |A_\lambda) = \sum_{k = 1}^{d_A} \alpha_k(\lambda)\, |P_{2k-1})\,, \label{AGP Series}
\end{align}
and Eq. \eqref{AGP_commutator_repn} in the matrix form satisfies
\begin{align}
    L^2\ket{A_\lambda} = -iL\ket{\partial_\lambda H}
\end{align}
with the superoperator $L_{mn} = (Q_{2m-1}|\mathcal{L}_\lambda|P_{2n-1})$. The operators can be represented as vectors in the Krylov basis as
\begin{align}
    &|A_\lambda) \longrightarrow \ket{A}= (\alpha_1, \alpha_2, \dots, \alpha_{d_A})^T; \;\; \alpha_k = (Q_{2k-1}|A_\lambda) \nonumber\\
    &|\partial_\lambda H) \longrightarrow \ket{\partial_\lambda H} = (1,0,\dots,0)
\end{align}
Thus, in a more explicit form, finding the AGP reduces to solving the matrix equation 
\begin{widetext}
    \begin{align}
\begin{pmatrix}
a_1^2 + b_1 c_1 + b_2c_2 & b_2 b_3  & \cdots & 0 \\
c_2 c_3 & a_3^2 + b_3 c_3 + c_4 b_4  & \cdots & 0 \\
0 & c_4 c_5  & \cdots & \vdots \\
\vdots & \vdots  & \ddots & b_{2d_A-1} b_{2d_A-2} \\
0 & 0  & c_{2d_A-1} c_{2d_A-2} & a_{2d_A-1}^2 + c_{2d_A} b_{2d_A} + c_{2d_A-1} b_{2d_A-1}
\end{pmatrix}
\begin{pmatrix}
    \alpha_1 \\ \alpha_2 \\ \vdots \\ \\ \alpha_{d_A} 
\end{pmatrix}
 = 
 \begin{pmatrix}
     -ic_1 \\ 0 \\\vdots \\  \\0
 \end{pmatrix}\,, \label{Mat_eq_biLanczos}
\end{align} 
\end{widetext}
where $d_A = K/2$ for even $K$, and $d_A = (K-1)/2$ for odd $K$. In the case of even $K$, the coefficients $c_{2d_A}$ and $b_{2d_A}$ do not exist and can be taken as zero. This is the generalization of the formulation in \cite{Takahashi_2024STA, Bhattacharjee_2023}, where the tridiagonal matrix had a simpler form with only one set of Lanczos coefficients $\{b_n\}$ due to Hermitian properties of $\mathcal{H}$. 

\medskip
We can also use the Krylov basis generated using Arnoldi iteration to find the AGP. The Arnoldi algorithm leads to a single basis $\vert K_k )$ in terms of which the generator $\mathcal{L}_\lambda$ can be written as
\begin{align}
    (K_{m}\vert\mathcal{L}_\lambda\vert K_n) = h_{m n}\Theta(n+1 - m)\,,\label{eq:arnoldi-alg-1}
\end{align}
where $\Theta$ is the Heaviside step function. For such a matrix in upper Hessenberg form, we can calculate the elements for its square as
 \begin{align}
     L^2_{jk} = \sum_{m = \max(0,j-1)}^{\min(K-1,k+1)} h_{jm}h_{mk}\,,
 \end{align}
 where $L_{jk} =(K_j|\mathcal{L}_\lambda|K_k)$, and we have used $h_{jk} \equiv h_{j,k}$ for brevity. Since we only need to take odd-indexed elements into account, the matrix $L^2$ reduces to the upper-Hessenberg form. As before, finding the AGP reduces to solving the matrix equation
 \begin{widetext}  
    \begin{align}
    \renewcommand{\arraystretch}{1.5} 
\begin{pmatrix}
h_{11}^2 + h_{10}h_{01} + h_{12}h_{21} & \sum_{m=0}^4h_{1m}h_{m3}  & \cdots & \sum_{m=0}^{2d_A} h_{1m}h_{m(2d_A-1)} \\
h_{32}h_{21} & h_{33}^2 + h_{32}h_{23} + h_{34}h_{43}  & \cdots &  \\
0 & h_{54}h_{43}  & \cdots & \vdots \\
\vdots & \vdots  & \ddots &  \\
0 & 0  &  & \sum_{m=2d_A-2}^{2d_A} h_{(2d_A-1)m}h_{m(2d_A-1)}
\end{pmatrix}
\begin{pmatrix}
    \alpha_1 \\ \alpha_2 \\ \vdots \\ \\ \alpha_{d_A} 
\end{pmatrix}
 = 
 \begin{pmatrix}
     -ih_{10} \\ 0 \\\vdots \\  \\0
 \end{pmatrix}\,. \label{Mat_eq_Arnoldi}
\end{align} 
\end{widetext}
Note that the terms $h_{2d_A,m}$ and $h_{m,2d_A}$ do not exist for even $K$ and should be set to zero.

As discussed earlier, the dimension of the Krylov space scales at most as $K \sim \mathcal{D}^2$ for a system with $\mathcal{D}$ dimensional Hilbert space. Thus, exploring the full Krylov space might not always be feasible for large many-body systems. The expression in Eq. \eqref{AGP Series} is exact if we include all the odd-indexed vectors in the Krylov space. However, this series can be truncated earlier for an approximate AGP
\begin{align}
      |A_\lambda^{(M)}) = \sum_{k=1}^{M<d_A}\alpha_k(\lambda) |P_{2k-1}) \label{Approx_AGP}
\end{align}
The effective dimension of the matrix equation to be solved will decrease accordingly. We will explore the effectiveness of this truncation in the following section.

\section{Variational Method for AGP}

Several works on counterdiabatic driving~\cite{Saberi2014,Sels2017,Kolodrubetz_2017,Claeys_2019} have introduced variational methods for evaluating the adiabatic gauge potential in complex Hermitian systems. These methods rely on finding an action, usually the Hilbert-Schmidt norm of an appropriate operator, whose extremization gives the adiabatic gauge potential $A_\lambda$. The formalism can be extended to non-Hermitian and Lindbladian systems by considering the AGP operator which is described as follows
\begin{align}
    \bra{L_m}A_\lambda\ket{R_n} = -i\frac{\bra{L_m}\partial_\lambda H \ket{R_n}}{E_m - E_n}\,,
\end{align}
where $m\neq n$ and the non-Hermitian Hamiltonian satisfies $H\ket{R_m} = E_m \ket{R_m}$ and $\bra{L_m}H = \bra{L_m}E_m$. This can be recast into the following form
\begin{align}
    \bra{L_m}i\partial_\lambda H + [H, A_\lambda]\ket{R_n} = 0\,.
\end{align}
Defining the operator $G_\lambda = i\partial_\lambda H + [H,A_\lambda]$, we can then compute $\bra{L_m}G_\lambda\ket{R_n}$, which gives the following result
\begin{align}
    \bra{L_m}G_\lambda\ket{R_n} = i\bra{L_m}\partial_\lambda H\ket{R_n}\delta_{m n} = i\partial_\lambda E_n\,,
\end{align}
which follows from application of Feynman-Hellmann theorem. Thus $G$ commutes with $H$, implying the constraint equation for $A_\lambda$
\begin{align}
    [H,G_\lambda] = [H,i\partial_\lambda H + [H,A_\lambda]] = 0\,.\label{eq:hGcomm1}
\end{align}
For non-Hermitian Hamiltonians $H^\dagger \neq H$ and so the same construction has to be prepared for $H^\dagger$ as well. This is represented by the equation
\begin{align}
    [H^\dagger, G^\dagger_\lambda] = -[H^\dagger, i \partial_\lambda H^\dagger + [H^\dagger_\lambda,A^\dagger_\lambda]] = 0\,.\label{eq:hGcomm2}
\end{align}
To write Eq.~\eqref{eq:hGcomm1}-\eqref{eq:hGcomm2} as saddle point equations of appropriate actions, we introduce the following \emph{biorthogonal} actions
\begin{align}
    S_{\lambda} = \overline{\Tr}(G^2_\lambda) \;\;,\;\; \tilde{S}_\lambda = \overline{\Tr}((G_\lambda^\dagger)^2)\,.
\end{align}
where the biorthogonal trace is defined as $\overline{\Tr}(M) = \sum_{m}\bra{L_m}M\ket{R_m}$. Extremizing $S_\lambda$ with respect to $A_\lambda$ gives the constraint equation $[H,G_\lambda] = 0$. Similarly the action $\tilde{S}_\lambda$ can also be extremized. Such actions are the natural quadratic bilinear forms associated with the biorthogonal space. In contrast to the Hermitian case, $S_\lambda$ is in general a complex number.
\subsection{Nested Commutators}
Given the variational action $G^2_\lambda$, an appropriate ansatz can be considered for $A_\lambda$, which can be then used to extremize $S_\lambda$ over all free parameters. A natural choice is to use a nested commutator expansion with a finite number of terms. Let us choose
\begin{align}
    A_{\lambda} = \sum_{k = 1}^{M}\alpha_k \mathcal{L}^{2k-1}(\partial_\lambda H)\,,\label{eq:agp-ansz}
\end{align}
where $M$ is the cutoff. Evaluating $G^2_\lambda$ using this gives us
\begin{align}
    G^2_\lambda &= \left(i\partial_\lambda H + \sum_{k = 1}^{M} \alpha_k\mathcal{L}^{2k}(\partial_\lambda H)\right)^2\notag\\
    &= -(\partial_\lambda H)^2 + i \sum_{k = 1}^{M}\alpha_k \{\mathcal{L}^{2k}(\partial_\lambda H),\partial_\lambda H\} \notag\\
    &\quad+ \sum_{k,k' = 1}^{M}\alpha_{k}\alpha_{k'}\mathcal{L}^{2k}(\partial_\lambda H)\mathcal{L}^{2k'}(\partial_\lambda H)\,.
\end{align}
The action can be evaluated as $\overline{\Tr}(G^2_\lambda)$, using which we can write extremization conditions as
\begin{align}
    \sum_{k'}\alpha_{k'}\overline{\Tr}(\mathcal{L}^{2k'}(\partial_\lambda H)\mathcal{L}^{2k}(\partial_\lambda H))
    = i\,\overline{\Tr}\!\left(\mathcal{L}^{2k}(\partial_\lambda H)\partial_\lambda H\right).
\end{align}
The sum on the LHS involves $M$ nested commutators. Since $\mathcal{L}^{2k}(\partial_\lambda H)$ do not satisfy any orthonormality relations, there is no natural truncation for Eq.~\eqref{eq:agp-ansz}. Convergence of the solution $\alpha_k$ is therefore slow in this basis. In the following section, we demonstrate that replacing the ansatz in Eq.~\eqref{eq:agp-ansz} by an orthonormal basis (specifically, the Krylov basis) provides a natural truncated series which leads to a finite-dimensional set of linear equations whose solution is the AGP.
\subsection{Krylov Ansatz}
The variational action $S_\lambda$ can be evaluated with the ansatz arising from the Krylov expansion (both via bi-Lanczos and via Arnoldi) with arbitrary coefficients. Let us consider the operator $A_\lambda = \sum_{k = 1}^{d_A}\alpha_k K_k$ described in Eq.~\eqref{eq:arnoldi-alg-1}. We use the operator notation for this section; the extension to state notation is straightforward. In the operator notation, the commutator $[H,K_n]$ can be written as
\begin{align}
    [H,K_n] = \sum_{j = 0}^{n+1}h_{jn}K_{j}\,.
\end{align}
In this notation, the matrix $G_\lambda$ can be written as 
\begin{align}
    G_\lambda = i b_0 K_0 + \sum_{k = 0}^{d_A}\alpha_{k}\sum_{j=0}^{k+1}h_{j k}K_j\,,\label{eq:g-lambda}
\end{align}
where $b_0 = \vert\vert\partial_\lambda H \vert\vert$. This can be simplified further by ordering $G_\lambda$ in terms of the coefficients of the vectors $K_j$. This can be written as 
\begin{align}
    G_\lambda = \sum_{j=0}^{d_A}G_j K_j\;,\;G_j = i b_0 \delta_{j,0} + \sum_{k=j-1}^{d_A}h_{jk}\alpha_k\,,
\end{align}
where we use the convention that $h_{jk} = 0\,\,\forall\,j,k < 0$. A key difference that must be emphasized here is that the norm, under which the action $S_\lambda$ is defined, is different from the norm under which both the Krylov algorithms are presented. The orthonormality condition states that $\Tr(K^\dagger_m K_n) = \delta_{m n}$, while the norm that appears in $S_\lambda$ is $\Tr(K_m K_n)$. Let us denote this quantity by the symmetric matrix $T_{m n} = \Tr(K_m K_n) = T_{n m}$. Using Eq.~\eqref{eq:g-lambda}, the action $S_\lambda$ can be written as $S_\lambda = \sum_{m,n}G_m T_{m n} G_n$. The variation of $G_m$ with respect to $\alpha_k$ is $h_{m k}$. With this one can write the extremization condition as
\begin{align}
    \frac{\delta S_\lambda}{\delta \alpha_k} = 2\sum_{m,n}G_n T_{n m}h_{m k} = 0\,.
\end{align}
Taking all $k$, this is equivalent to the matrix relation
\begin{align}
    \mathbf{h}^T\mathbf{T} \vec{G} = \mathbf{0}\,.
\end{align}
This can be simplified further by noting that $\mathbf{h}^T\mathbf{T} = -\mathbf{T}\mathbf{h}$, which follows from the Arnoldi expansion Eq.~\eqref{eq:arnoldi-alg-1} and the symmetry of $\mathbf{T}$ under transpose. The effective equation to solve is therefore $\mathbf{h}\cdot\vec{G} = \mathbf{0}$. This further simplifies to
\begin{align}
    \mathbf{h}^2\cdot \vec{\alpha} = -ib_0 (h_{00}\vec{e}_0 + h_{10}\vec{e}_1)\,.\label{eq:var-eq-arnoldi}
\end{align}
under the assumption that $\det(\mathbf{T}) \neq 0$. The only non-vanishing contribution comes from the odd vectors $K_{2k-1}$ (for the $\mathcal{PT}-$symmetric systems considered here), since for the even coefficients $\alpha_{2k}$ satisfy an equation of the form Eq.~\eqref{Mat_eq_Arnoldi} with a vector of $0$'s on the RHS. Thus the equation is solved by setting all $\alpha_{2k} = 0$. Thus Eq.~\eqref{eq:var-eq-arnoldi} agrees exactly with Eq.~\eqref{Mat_eq_Arnoldi} for $h_{00} = 0$ (true for $\mathcal{PT}-$symmetric systems), demonstrating that the Arnoldi basis extremizes the variational action.

The next step is to consider the bi-Lanczos approach. The ansatz that is chosen for the AGP is $A_\lambda = \sum_{k}\alpha_k P_k$, where $P_k$ are the right Krylov operators as described in Eq.~\eqref{eq:bilanczos-alg-1}. The bi-Lanczos algorithm then allows one to write
\begin{align}
    G_\lambda &= \sum_{j = 0}^{d_A}G_j P_j\;,\;\notag\\ G_j &= i b_0\delta_{j 0} + \alpha_{j-1}c_{j} + \alpha_j a_j + \alpha_{j + 1}b_{j+1}\,,
\end{align}
where the constraint is that $c_{j \leq 0} = 0$ and $b_{j \geq d_A} = 0$. Correspondingly the derivative is $\partial G_j/\partial \alpha_k = L_{k j} = \delta_{k,j-1}c_{k} + \delta_{k j}a_k + \delta_{k,j+1}b_k$. Following the same analysis as for the Arnoldi method, we can write
\begin{align}
    \mathbf{L}^2\cdot \vec{\alpha} = -i b_0 c_1 \vec{e}_1\,,
\end{align}
where the non-vanishing contribution arises from the odd-indexed coefficients.

\section{Examples}
\subsection{Application to decaying two-level atoms}
We will apply the above formulation to a few physical systems. As a proof of concept, we can calculate AGP for a decaying two-level atom where the spontaneous decay is modulated by a chirped laser with time-dependent frequency. Under certain assumptions, including electric dipole approximation, laser-adapted interaction picture, and the rotating wave approximation, we can write the Hamiltonian as \cite{Ibanez2011, Ibanez2011ChirpedPulses}
\begin{align}
    H(t) &=  \frac{\hbar}{2}\begin{pmatrix}
        -\Delta(t) & \Omega(t) \nonumber \\ 
        \Omega(t) & \Delta(t) - i\Gamma(t)
    \end{pmatrix} \\
    &= \frac{\hbar}{2}\Big[\Omega(t) \sigma_x - \Delta(t)\sigma_z + \frac{i\Gamma}{2}(\sigma_z - \mathbb{I})\Big]\,,\label{eq:h-2lvl}
\end{align}
in the usual Pauli-$z$ basis. Here, $\Delta(t) = \omega_0 - \omega_i(t)$ captures the detuning from the atomic transition frequency with $\omega_i(t)$ as the time-dependent instantaneous field frequency, $\Omega(t)$ is the Rabi frequency, and $\Gamma(t)$ is the effective decay rate. We can change the above parameters as a function of time, which serves as the tuning parameter $\lambda$. The time derivative of the Hamiltonian can be written in the same basis as
\begin{align}
    \partial_tH(t) = \frac{\hbar}{2}\Big[\dot{\Omega}(t) \sigma_x - \dot{\Delta}(t)\sigma_z + \frac{i\dot{\Gamma}}{2}(\sigma_z - \mathbb{I})\Big]\,.
\end{align}
In the basis defined by the Pauli matrices and using the Frobenius inner product, we can represent the superoperator $\mathcal{L}$ and the operator $\partial_tH(t)$ as a matrix and a vector, respectively, which allows us to apply the explicit bi-Lanczos algorithm. The details of the calculation are given in Appendix \ref{Krylov basis}. 

We note that the Krylov basis is three-dimensional, which implies that there is only one term in the series expansion in Eq. \eqref{AGP Series}. We can calculate the coefficient $\alpha_1$ as
\begin{align}
    \alpha_1 = \frac{-ic_1}{b_1c_1 + b_2c_2} = \frac{\dot{\Omega}(\Delta - i\Gamma/2) - \Omega(\dot{\Delta} - i\dot{\Gamma}/2)}{(\Delta - i\Gamma/2)^2 + \Omega^2}\,,\label{eq:alpha-1}
\end{align}
which gives the AGP as
\begin{align}
    H_{\rm{cd}}(t) = \alpha_1 |P_1) = \frac{\hbar}{2}\begin{pmatrix}
        0 & i\alpha_1 \\
        -i\alpha_1 & 0
    \end{pmatrix}\,.
\end{align}
This is the same matrix calculated using exact diagonalization in \cite{Ibanez2011}. The non-Hermitian Hamiltonian has an exceptional point at $\Gamma(t) = 2\Omega(t)$ and $\Delta(t) = 0$. It is interesting to note that the AGP coefficient $\alpha_1$ also diverges at that point \cite{Ibanez2011}. 

For this $2-$level system, the variational approach can be applied to computing the AGP. The action $\Tr(G^2_t)$ is computed with the ansatz $A = a\sigma_{x} + b\sigma_{y} + c\sigma_{z} + d\mathbb{I}$. Extremizing $S_t$ gives a self-consistent set of equations for $\{a, c, d\}$ which are solved by the choice $a = c = d = 0$. The independent term comes from the equation for $b$, which is the result obtained in Eq.~\eqref{eq:alpha-1} as $\alpha_1$. 

\subsection{Application to a many-body system: the Hatano-Nelson Model}

The true advantage of writing AGP as a series of nested commutators lies in many-body systems, where exact diagonalization is not always feasible. As an example of a non-Hermitian many-body system, we will now calculate AGP in the interacting Hatano-Nelson model with a finite-time ramp of imaginary vector potential.

The Hatano-Nelson model describes fermions in the presence of an imaginary vector potential \cite{HatanoNelson1996, HatanoNelson1997}. Its Hamiltonian can be written as
\begin{eqnarray}
    H_{\rm HN}(t) &=& \frac{J}{2}\sum_{n=1}^{N-1} \left(e^{ah(t)}c_n^\dagger c_{n+1}+e^{-ah(t)}c_{n+1}^\dagger c_{n}\right) \nonumber \\ 
    & & +U \sum_{n=1}^{N-1} c_n^\dagger c_n c_{n+1}^\dagger c_{n+1}\,,
\end{eqnarray}
where $J>0$ captures uniform hopping, $h(t)$ denote time-dependent imaginary vector potential, $a$ is the lattice constant, and $U$ is the nearest neighbor interaction between the particles. There are $N$ lattice sites, and we consider an open boundary condition in a fixed particle sector ($N/2$ particles). This model was studied in \cite{Dupays2025} in the context of adiabaticity for a finite-time linear ramp $h(t) = h_0t/\tau$ for $0\leq t \leq \tau$. Using parameters such as excess energy, density imbalance and the Loschmidt echo, the system was shown to reach the adiabatic limit $\tau \to \infty$ with decay scaling as $\tau^{-1}$ \cite{Dupays2025}. For a fixed imaginary vector potential $h(t)$, it can be shown that the instantaneous spectrum remains unchanged and is equal to $h=0$ case for the open boundary condition \cite{DoraQuench2023}. This ensures the weak non-Hermiticity condition in  Eq. \eqref{weak non-Hermiticity} and validates our approach. 

\begin{figure}[ht]
    \centering
    \includegraphics[width=0.9\linewidth]{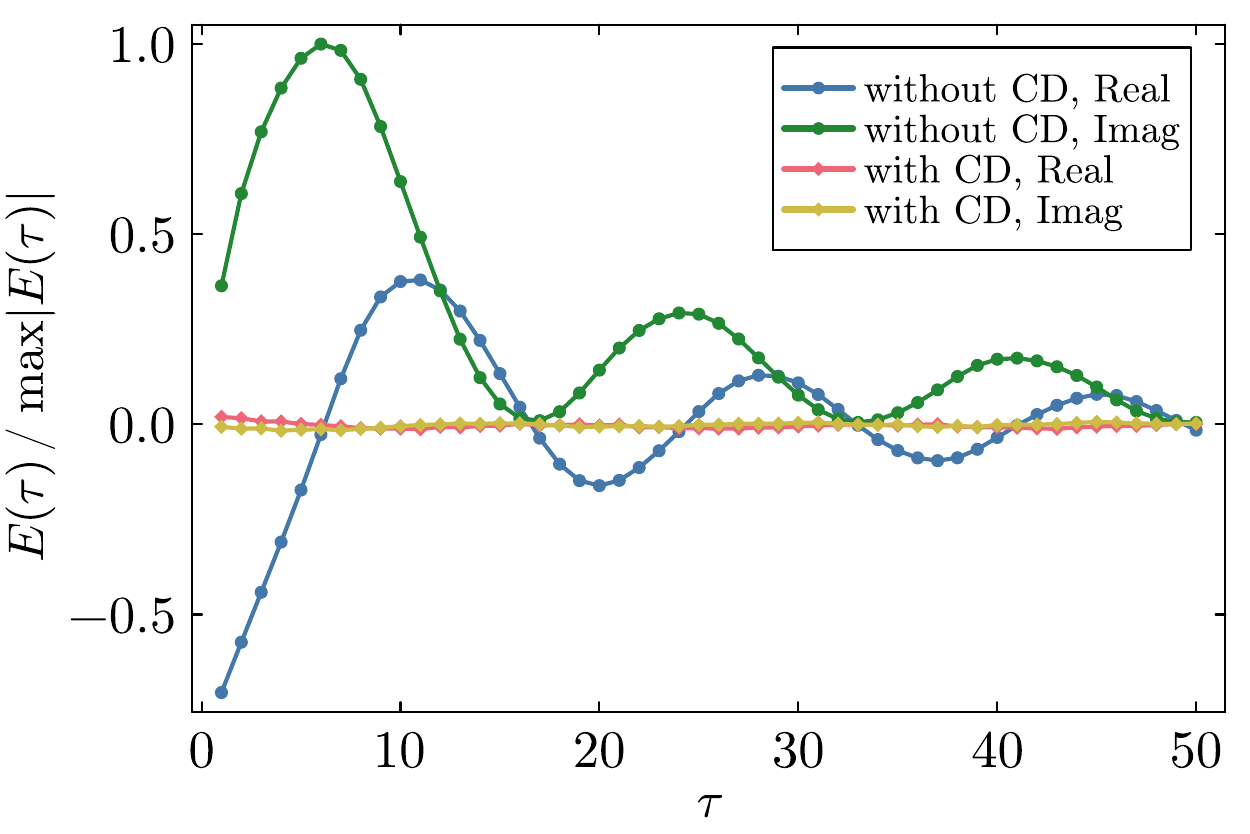}
    \caption{Excess energy $E(\tau)$ at the end of different ramp time $\tau$ for the interacting Hatano-Nelson model for $N=10$ with and without the counterdiabatic driving term ($J = 1.0, U =1.0, a = 1.0$, and $h_0 = 0.1$). AGP was truncated at $M=80$, where the exact AGP would require $M=31626$ terms ($\lfloor K/2 \rfloor$ for $\mathcal{D} = 252)$.}
    \label{fig: Excess Energy}
\end{figure}

\begin{figure}[ht]
    \centering
    \includegraphics[width=0.9\linewidth]{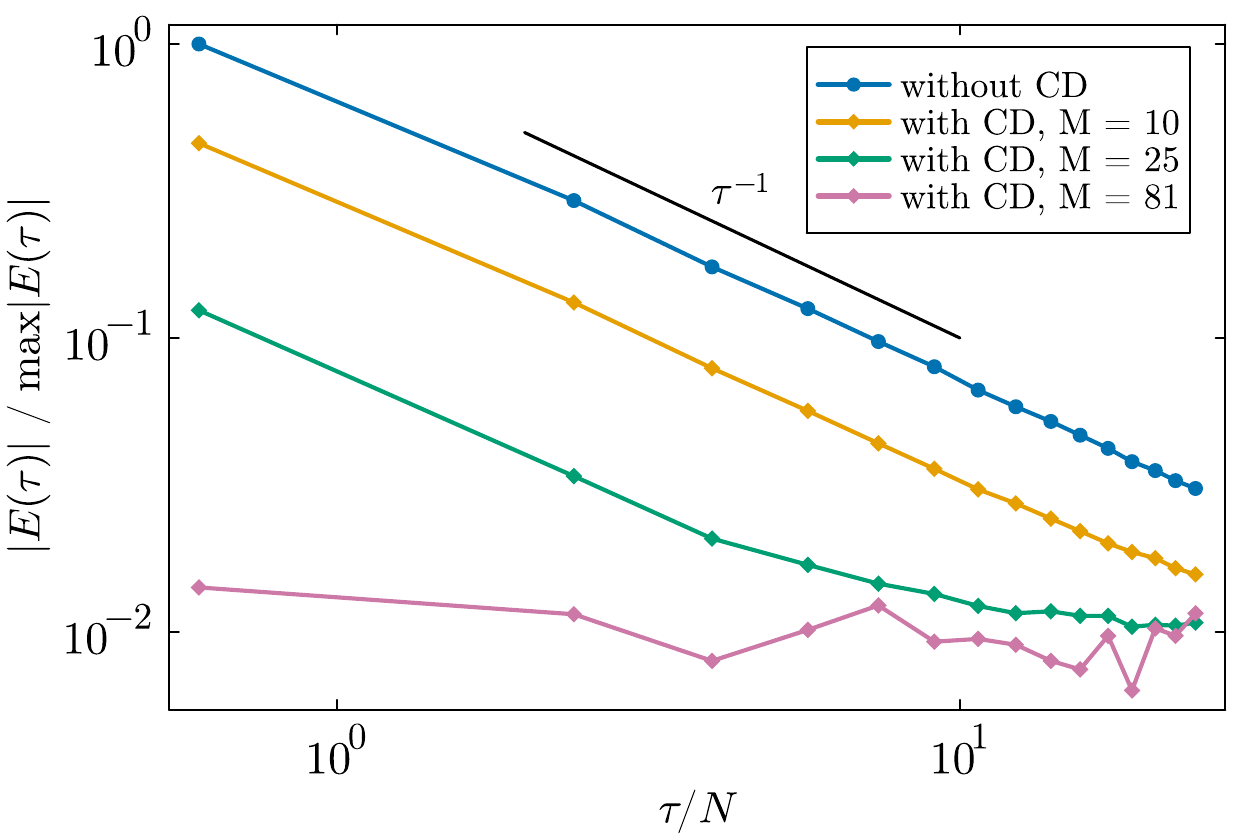}
    \caption{Absolute value of the excess energy $|E(\tau)|$ plotted at the maxima points during the oscillation for different approximate AGPs ($N=10, \mathcal{D} = 252$). The exact AGP requires $M = 31626$ terms ($\lfloor K/2 \rfloor$). }
    \label{fig: Excess Energy Peak}
\end{figure}

We will consider the non-Hermitian generalization of excess energy, which captures the difference between final mean energy and adiabatic mean energy. It is defined as
\begin{align}
    E(t) = \frac{\bra{\Psi(t)}H(t) \ket{\Psi(t)}}{\braket{\Psi(t)}{\Psi(t)}} - E_0\,,
\end{align}
where $E_0$ is the energy of the ground state $\ket{\Psi(0)}$ at $t=0$. The numerical result for a finite-time linear ramp is shown in Fig. \ref{fig: Excess Energy}, with and without the counterdiabatic driving term, where we consider the excess energy at the end of the ramp. The excess energy is clearly suppressed compared to the natural evolution when the counterdiabatic driving terms are added. We also drive the system with different approximate AGPs \eqref{Approx_AGP}, which are constructed with a finite number of Krylov basis vectors. The results are shown in Fig. \ref{fig: Excess Energy Peak}, which shows clear and fast convergence towards the effect of full counterdiabatic driving.

\subsection{Application to a $\mathcal{PT}-$ symmetric system: the isotropic Heisenberg spin chain}

An important class of non-Hermitian systems that have real spectrum is parity-time symmetric (or $\mathcal{PT}-$symmetric) systems \cite{Bender_PT}. They describe physical systems that are borderline between an open and a closed system. The effect of environment on such systems is restricted such that there are effectively no loss or gain. Hamiltonians describing $\mathcal{PT}-$ symmetric systems have either real eigenvalues (corresponding to the $\mathcal{PT}-$unbroken phase when the eigenvectors are also $\mathcal{PT}-$ symmetric) or come in conjugate pairs (corresponding to the $\mathcal{PT}-$broken phase when eigenvectors break the $\mathcal{PT}-$ symmetry) (see \cite{Ashida_NHPhysics} for a review). We will look at the isotropic Heisenberg spin chain as an example of such a system, which was exactly solved using Bethe ansatz in \cite{Kattel_PT_2023}.

Consider a Heisenberg spin chain with complex magnetic fields applied to the edges in $z-$direction. An effective Hamiltonian can be written as
\begin{align}
    H = \sum_{j=1}^{N-1} \sum_{\alpha = \{x,y,z\}} \sigma_j^\alpha \sigma_{j+1}^\alpha + \frac{1}{\xi + i\chi}\sigma_1^z + \frac{1}{\xi - i\chi}\sigma_N^z\,,
\end{align}
for $N$ spin sites using the set of Pauli matrices $\{\sigma^\alpha\}$. The action of parity and time-reversal are $P\sigma_j^\alpha P = \sigma_{N+1-j}^\alpha$ and $TiT = -i$ respectively. Under the action of the joint $PT$ operator, the Hamiltonian is symmetric. It can be shown analytically that the Hamiltonian has $\mathcal{PT}-$unbroken phase for $|\xi|>1/2$ and a mixed broken/unbroken phase for $|\xi|<1/2$ \cite{Kattel_PT_2023}. The ground states in all phases are non-degenerate for odd numbers of spin sites, and we can apply our formalism of generating the AGP while tuning the phase parameter $\xi$.

We can use the AGP norm to benchmark the convergence of the approximate AGP. The AGP norm is defined as the Frobenius norm of the AGP operator and captures the magnitude of the counterdiabatic term needed for fast driving. It can be generalized to non-Hermitian Hamiltonians as 
\begin{align}
    ||A_\lambda||^2 = (A_\lambda|A_\lambda) = \sum_{m\neq n} \frac{|\bra{L_m}\partial_\lambda H\ket{R_n}|^2}{|E_m-E_n|^2}\,, \label{exact_AGP_norm}
\end{align}
and it takes a simple form in the orthonormal Krylov basis
\begin{align}
    ||A_\lambda||^2 = \sum_{k=1}^{d_A}|\alpha_k|^2\,.
\end{align}
Here, elements $\{\alpha_k\}$ are the solution to matrix equations \eqref{Mat_eq_biLanczos} or \eqref{Mat_eq_Arnoldi}. To benchmark the approximate AGP defined in \eqref{Approx_AGP}, we can use the approximate AGP norm
\begin{align}
    ||A^{(M)}_\lambda||^2 \equiv (A_\lambda^{(M)}|A_\lambda^{(M)}) = \sum_{k=1}^{M<d_A}|\alpha_k|^2\,. \label{app. AGP norm}
\end{align}
It should approach the exact AGP norm when all the odd-indexed vectors from the Krylov basis are used.

\begin{figure}[ht]
    \centering
    \includegraphics[width=1.0\linewidth]{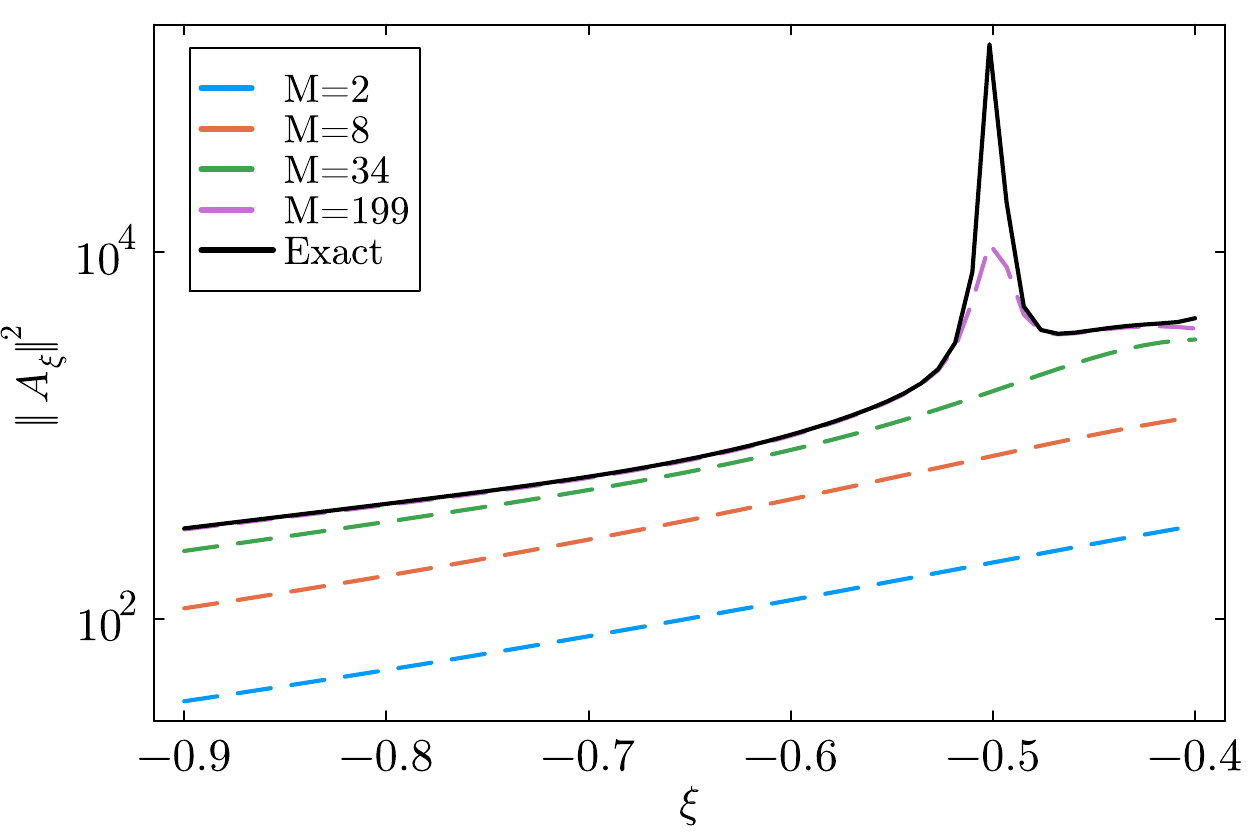}
    \caption{AGP norm for the $\mathcal{PT}-$symmetric Heisenberg chain $(N=9, \mathcal{D} = 512)$ using finite number of Krylov basis vectors across the phase parameter $\xi$. Exact AGP would require $M=130816$ terms ($\lfloor K/2 \rfloor$).}
    \label{fig: AGP norm}
\end{figure}

Figure \ref{fig: AGP norm} shows the AGP norm (Eq. \eqref{app. AGP norm}) for different  approximate AGPs constructed using first $M$ odd-indexed Krylov basis vectors. For the calculation, we fix $\chi \approx 10^{-7}$ so that the Hamiltonian is non-Hermitian, but the spectrum is real up to the machine precision. We can see that it rapidly approaches the exact AGP norm in Eq. \eqref{exact_AGP_norm}. Moreover, it also captures the phase transition at $\xi = 1/2$. 

We emphasize that the norm of the AGP is used here to determine the truncation of the Krylov expansion. This does not preclude the existence of efficient AGP approximations that effectively achieve CD and yet have a norm much smaller than the exact AGP, as it occurs in the quantum brachistochrone counterdiabatic driving of models with spin-glass bottlenecks \cite{Grabarits2026gaps}.

\subsection{Exactly solvable $\mathcal{PT}-$ symmetric chain: non-Hermitian transverse field Ising model}
An example where the full mechanism of the Krylov approach can be seen to emerge is a non-Hermitian spin chain which is analytically solvable. We consider the non-Hermitian transverse field Ising model (NH-TFIM), which is described by the Hamiltonian~\cite{Lu2024msnybody,Yang2022hidden}
\begin{align}
    H = -J\sum_{j = 1}^{L}\sigma^{x}_j \sigma^{x}_{j+1} + \sum_{j=1}^{L}h(\sigma^{z}_j + i\gamma\sigma^{y}_j)\,.
\end{align}
Here, $\sigma^\alpha_{j}$ are the three Pauli matrices for $\alpha = x, y, z$ at site $j$. Periodic boundary conditions are imposed. The model is known to be $\mathcal{PT}$ symmetric, with a $\mathcal{PT}-$unbroken phase for $\gamma < 1$ and a $\mathcal{PT}-$broken phase for $\gamma > 1$. This model can be mapped to a transverse field Ising model by the Schrieffer-Wolff transformation $\tau^{z}_j = e^{\frac{\beta}{2}\sigma^{x}_j}\sigma^{z}_j e^{-\frac{\beta}{2}\sigma^{x}_j}$, where $\beta = \frac{1}{2}\ln\left(\frac{1+\gamma}{1-\gamma}\right)$. This transforms the Hamiltonian into
\begin{align}
    H = -J \sum_{j=1}^{L}\tau^{x}_{j}\tau^{x}_{j+1} + h \sqrt{1-\gamma^2} \sum_{j = 1}^{L}\tau^{z}_j\,.\label{eq:h-tfim}
\end{align}
This model has an Ising transition at $\gamma_{c} = \sqrt{1 - \frac{J^2}{h^2}}$~\cite{Sun2021biorthogonal,Yang2022hidden}. This Hamiltonian can be cast into the uncoupled $2-$level system form (via a Jordan-Wigner transformation) $H = \sum_{k}\psi^\dagger_k H_k \psi_k$ with $H_k = (-2 J \cos k - 2 g)\sigma^z + (2 J \sin k)\sigma^y$, and the energy spectrum given by $E_k = 2\sqrt{(-J \cos k - g)^2 + (J \sin k)^2}$. Here, $g = h \sqrt{1 - \gamma^2}$ and $\psi_k = (c_k, c^\dagger_{-k})$. The Krylov basis for the Hermitian TFIM has been constructed exactly~\cite{Takahashi_2024STA}. The fermion bilinear form of the NH-TFIM suggests that the following operator can be used to construct a Krylov basis
\begin{align}
    W_{k} = \frac{1}{\sqrt{2}} \sum_{j=1}^{L} \left(\tau^{x}_jZ_{j,k}\tau^{y}_{j+k} + \tau^{y}_jZ_{j,k}\tau^{x}_{j+k}\right)\,,\label{eq:W_k_full}
\end{align}
where we use $Z_{j,k} \equiv \prod_{l = j+1}^{j + k -1}\tau^{z}_l$, with $Z_{j,1} = \mathbb{I}$. As we discuss in detail in Appendix~\ref{Krylov nhtfim}, the individual components in the sum in Eq.~\eqref{eq:W_k_full} also appear in the Krylov algorithm, along with the total magnetization $M = \sum_{j = 1}^{L}\tau^{z}_j$. These can be denoted by $V^{x,y}_k = \sum_{j=1}^{L}\tau^{x,y}_jZ_{j,k}\tau^{x,y}_{j+k}$. The fermion bilinear operators satisfy $\vert\vert M \vert\vert^2= \vert\vert V^{x,y}_k\vert\vert^2 = \vert\vert W_k\vert\vert^2 = L$. The Krylov basis can be constructed by choosing $g(t) = h\sqrt{1-\gamma^2(t)}$ as the driving parameter and $\partial_g H = M$. 
This can be now used to construct the Krylov basis, starting from the initial operator $M$ and recursively applying the Arnoldi algorithm. We consider the 2 phases separately: $\gamma < 1$ and $\gamma > 1$. 
\subsubsection{PT unbroken: $\gamma < 1$}
In this regime, the Hamiltonian in Eq.~\eqref{eq:h-tfim} is Hermitian. Since the initial operator $M$ is also Hermitian, the Arnoldi/bi-Lanczos algorithms become the usual Lanczos algorithm, providing the same results as derived in~\cite{Takahashi_2024STA}. We mention the same below.

The Krylov vectors are $K_n = (-1)^n i W_n/\sqrt{L}$ and the diagonal Krylov coefficients are $a_n = 0$. The non-zero off-diagonal coefficients are the tridiagonal ones $b_k$, which satisfy the constraint 
\begin{align}
    b^2_{2k-1} + b^2_{2k} &= 16 J^2 \left( 1 + \frac{g^2}{J^2}\right)\,,\label{eq:bn-const-1}\\
    b_{2k}b_{2k+1} &= -16 J g\,.\label{eq:bn-const-2}
\end{align}
This makes the matrix in Eq.~\eqref{Mat_eq_biLanczos} a tridiagonal Toeplitz matrix, with all diagonal elements equal to $16 J^2 \left( 1 + \frac{g^2}{J^2}\right)$ and all off-diagonal elements equal to $-16 J g$. The equation can therefore be solved using Fourier inversion. The CD term can be expanded in terms of $K_n$ to give $A_\lambda = i\sum_{n=1}^{L/2}(-1)^n\alpha_n W_n/\sqrt{L}$. The Krylov dimension for chosen initial operator $M$ is $L/2$. 

\subsubsection{PT broken : $\gamma > 1$}
This is the sector where the analysis deviates from the usual TFIM. For this, we employ the bi-Lanczos algorithm, where we note that the right Krylov vectors are propagated by $\mathcal{L}_g = [H,\cdot]$ and the left Krylov vectors are propagated by $\mathcal{L}^\dagger_g = [H^\dagger,\cdot]$. For $\gamma > 1$, we can write Eq.~\eqref{eq:h-tfim} as $H = H_J - i\mu M$ where $\mu = h\sqrt{\gamma^2-1} \in \mathbb{R}$ and $H_J = -J\sum_{i}\tau^{x}_{i}\tau^{x}_{i+1}$. It follows that $\mathcal{L}_g^\dagger = \mathcal{L}_g\vert_{\mu\to-\mu}$.  

Starting with the initial operator $\hat{O} = M/\sqrt{L}$ and applying the bi-Lanczos algorithm from Appendix ~\ref{sec:bilanczos}, we construct the tridiagonal representation of the Liouvillian. This is given in terms of the upper and lower off-diagonal elements $c_n, b_n$ and the diagonal components $a_n$. As we demonstrate in Appendix~\ref{Krylov nhtfim}, the diagonal components vanish, while the off-diagonal terms satisfy the relation
\begin{align}
    b_{2k-1}c_{2k-1} + b_{2k}c_{2k} &= 16 J^2 \left(1+ \frac{g^2}{J^2}\right)\,,\\
    b_{2k}b_{2k + 1} = c_{2k+2}c_{2k+3} &= -16 J g\,.
\end{align}
These constraint relations are the generalisation of Eqs.~\eqref{eq:bn-const-1}-\eqref{eq:bn-const-2} to the bi-Lanczos basis. These are valid in both $\mathcal{PT}$-broken and unbroken phases. The resulting matrix on the LHS of Eq.~\eqref{Mat_eq_biLanczos} is a tridiagonal Toeplitz matrix, which takes the form $\mathbf{T}\cdot\vec{\alpha} = -i c_1 \vec{e}_1$. Each $\alpha_k$ is simply $\alpha_k = -i c_1 (\mathbf{T}^{-1})_{k1}$, with $c_1 = b_1 = 2\sqrt{2}J$. The eigenvalues of a $\mathbf{T}$ are given by
\begin{align}
    &\lambda_k = a + 2d \cos(\theta_k),\notag\\
    &\text{with}\;\;a = 16(J^2 + g^2)\,,\,d = -16 J g\,,\,\theta_k = \frac{\pi k}{d_A + 1},
\end{align}
where $d_A$ is the dimension of $\mathbf{T}$, the inverse matrix elements can be written as 
\begin{align}
    \left(\mathbf{T}^{-1}\right)_{m n} = \frac{2}{d_A + 1}\sum_{k = 1}^{d_A}\frac{\sin(m\theta_k)\sin(n\theta_k)}{\lambda_k}\,.
\end{align}
This matrix element can be computed recursively in terms of minors of $\mathbf{T}^{-1}$ using Usmani's formula~\cite{usmani1994inversion}. This gives the form of the inverse matrix element as
\begin{align}
    \left(\mathbf{T}^{-1}\right)_{k 1} = (-1)^{k+1}d^{k-1}\frac{D_{d_A - k}}{D_{d_A}}\,,
\end{align}
where the $n^\text{th}$ minor is denoted by $D_n$. The minors obey the recursion relation
\begin{align}
    D_n = a D_{n-1} - d^2 D_{n-2}\;, \;D_0 = 1\,,D_1 = a\,,
\end{align}
which is then solved by introducing the parameter $\nu$ defined via $\cosh(\nu) = \frac{a}{2 d}$, equivalently $\nu = \ln (J/g)$. This is complex in the $\mathcal{PT}-$broken regime $\gamma > 1$. The solution for $D_n$ is given by
\begin{align}
    D_n = (16 J g)^n \frac{\sinh((n + 1)\nu)}{\sinh\nu}\,.
\end{align}
Thus the final expression for $\alpha_k$ is given by
\begin{align}
    \alpha_k =  \frac{-i}{4\sqrt{2}g}\frac{\sinh\left((d_A + 1 - k)\cosh^{-1}\left(\frac{J^2 + g^2}{2J g}\right)\right)}{\sinh\left((d_A + 1)\cosh^{-1}\left(\frac{J^2 + g^2}{2J g}\right)\right)}.
\end{align}
The norm of the Adiabatic Gauge Potential follows from the sesquilinear sum
$\Vert A_\lambda\Vert^2 = \sum_{k=1}^{d_A}\vert\alpha_k\vert^2$. Writing
$\nu = \nu_R + i\nu_I = \ln(J/g)$ and using
$\vert\sinh z\vert^2 = \sinh^2(\mathrm{Re}\,z) + \sin^2(\mathrm{Im}\,z)$,
the same-frame contributions combine into the manifestly real, non-negative form
\begin{align}
    \Vert A_\lambda\Vert^2 = \frac{1}{128\,\vert g\vert^2}\,
    \frac{\dfrac{\sinh\!\big((2d_A + 1)\nu_R\big)}{\sinh\nu_R}
        - \dfrac{\sin\!\big((2d_A + 1)\nu_I\big)}{\sin\nu_I}}
    {\sinh^2\!\big((d_A + 1)\nu_R\big) + \sin^2\!\big((d_A + 1)\nu_I\big)}\,.
    \label{eq:agp-norm-nhtfim}
\end{align}

In the unbroken ($\mathcal{PT}$-symmetric) phase $\gamma < 1$, $g$ is real and
$\nu_I = 0$; the second term in the numerator reduces to $2d_A + 1$ and the
expression collapses to
\begin{align}
    \Vert A_\lambda\Vert^2 = &\frac{1}{128\,g^2}
    \left(\frac{\sinh\!\big((2d_A + 1)\nu\big)}{\sinh\nu} - (2d_A + 1)\right)\notag\\&\times
    \csch^2\!\big((d_A + 1)\nu\big)\,.
    \label{eq:agp-norm-unbroken}
\end{align}

In the broken phase $\gamma > 1$, $g = i\mu$ with $\mu = h\sqrt{\gamma^2 - 1}$,
so that $\nu_R = \ln(J/\mu)$ and $\nu_I = -\pi/2$. For even $d_A$ the trigonometric
factors reduce to parities and the norm simplifies to
\begin{align}
    \Vert A_\lambda\Vert^2\big\vert_{\gamma > 1}
    = \frac{1}{64\,\mu^2}\,
    \frac{\sinh\!\big(d_A\,\nu_R\big)}
    {\sinh(\nu_R)\,\cosh\!\big((d_A + 1)\nu_R\big)}\,,
    \label{eq:agp-norm-broken}
\end{align}
where $\nu_R = \ln(J/\mu)$, and the norm remains positive and finite across the exceptional-point line $\mu = J$,
where it takes the value $d_A/(64 J^2)$.

\begin{figure}[t]
    \centering
    \includegraphics[width=1.0
    \linewidth]{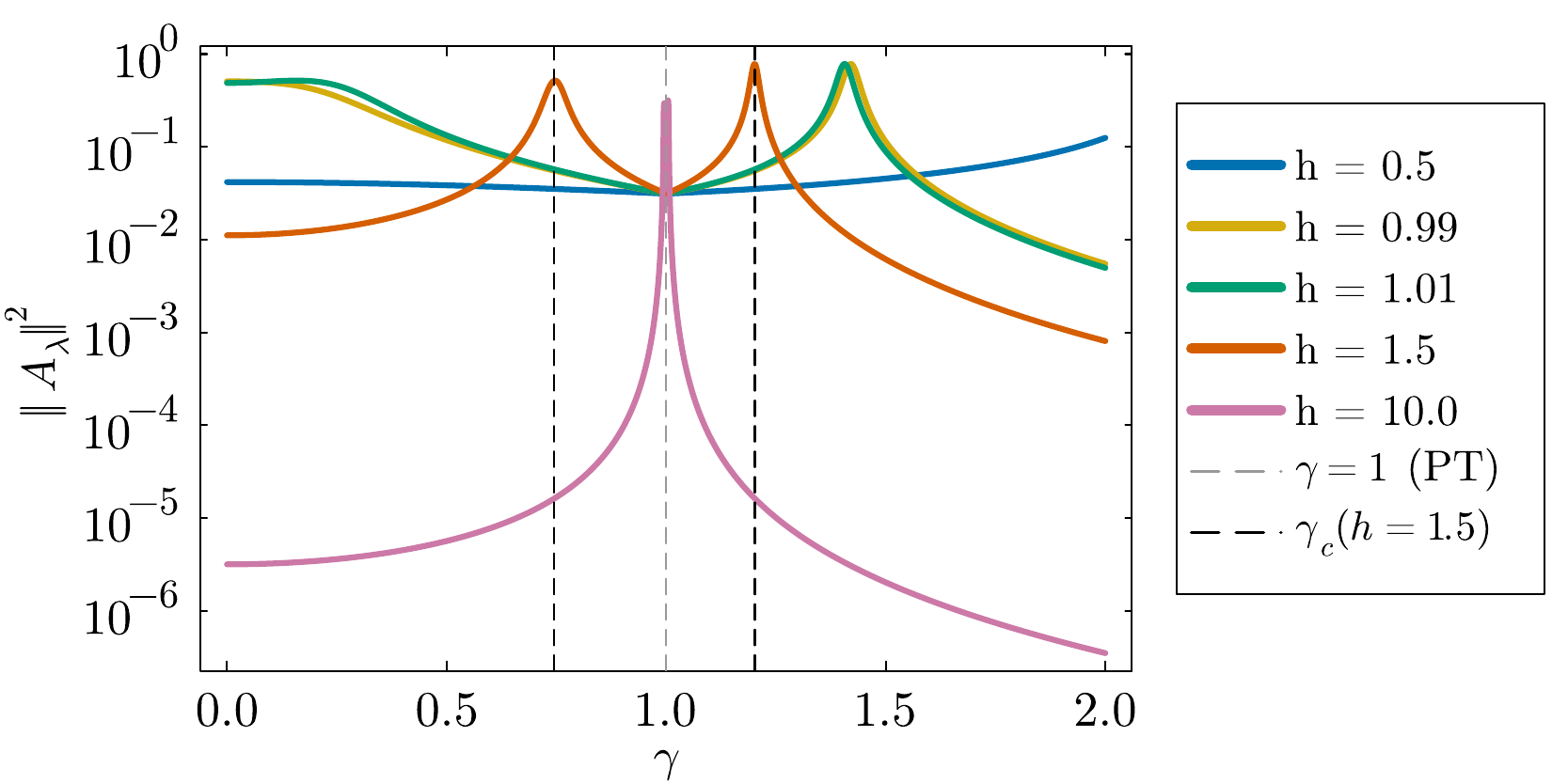}
    \caption{Norm of the adiabatic gauge potential for the Non-Hermitian TFIM, as a function of the parameter $\gamma$. The Ising transition is shown for $h = 1.5$. The Krylov space dimension is taken to be $d_A = 50$ for this result.}
    \label{fig:AGP_NHTFIM}
\end{figure}

The norm of the AGP captures the $\mathcal{PT}-$symmetry breaking transition at $\gamma = 1$, along with the Ising transition at $\gamma_c = \sqrt{1-(J/h)^2}$. This is presented in Fig.~\ref{fig:AGP_NHTFIM}, where there is a kink at $\gamma = 1$ for small $h$ and a peak at the Ising transition $\gamma_c$. This transition corresponds to the limit $\nu \to 0$. In this limit, the scaling of $\vert\vert A_\lambda \vert\vert^2$ can be computed from Eq.~\eqref{eq:agp-norm-nhtfim}, which gives us
\begin{align}
    \vert\vert A_\lambda\vert\vert^2\Big\vert_{\gamma < 1} \underset{\nu \to 0}{\to} \frac{L}{192 J^2}\,.
\end{align}

The second feature in the AGP norm occurs at
$\gamma_c = \sqrt{1 + J^2/h^2}$, where $\mu = h\sqrt{\gamma_c^2 - 1} = J$ and
hence $\nu_R = \ln(J/\mu) = 0$. This is precisely the exceptional-point line of
the non-Hermitian model, at which the biorthogonal frame becomes singular. The
factor $\sinh\nu_R$ in the denominator of $\eqref{eq:agp-norm-broken}$ signals
the associated coalescence, but the numerator $\sinh(d_A\nu_R)$ vanishes at the
same rate, so the norm stays finite and the divergence is averted. Expanding
$\eqref{eq:agp-norm-broken}$ about $\nu_R \to 0$ with
$\sinh(d_A\nu_R) \simeq d_A\nu_R$, $\sinh\nu_R \simeq \nu_R$, and
$\cosh((d_A+1)\nu_R) \simeq 1$ gives
\begin{align}
    \Vert A_\lambda\Vert^2 \Big\vert_{\gamma > 1} \underset{\nu_R \to 0}{\to}
    \frac{L}{128\,J^2}\,,
\end{align}
so at the exceptional point the AGP norm grows \emph{linearly} in system size,
$\Vert A_\lambda\Vert^2 \sim L$. This extensive scaling marks $\gamma_c$ as a
line of enhanced adiabatic susceptibility distinct from the $\mathcal{PT}$
transition at $\gamma = 1$, reflecting the diverging cost of counterdiabatic
driving as the eigenvectors coalesce.

At the $\mathcal{PT}$ transition $\gamma = 1$ one has $g = h\sqrt{1-\gamma^2} \to 0$,
so that $\nu = \ln(J/g) \to +\infty$. Unlike the exceptional point, this is a limit
in which the hyperbolic factors of \eqref{eq:agp-norm-unbroken} grow without bound,
but the exponentials cancel: using
$\sinh\!\big((2d_A+1)\nu\big)/\sinh\nu \simeq e^{2d_A\nu}$ and
$\operatorname{csch}^2\!\big((d_A+1)\nu\big) \simeq 4\,e^{-2(d_A+1)\nu}$, the
$L$-dependent exponents cancel identically and the surviving prefactor
$4\,e^{-2\nu}/(128\,g^2) = (g^2/J^2)/(32\,g^2)$ is finite, giving
\begin{align}
    \Vert A_\lambda\Vert^2\big\vert_{\gamma = 1}
    = \frac{1}{32\,J^2}\,,\label{eq:agpnorm-scaling-PT}
\end{align}
with corrections of order $e^{-2\nu} = (g/J)^2$. The same value is recovered from
the broken-phase form \eqref{eq:agp-norm-broken} as $\mu \to 0$, so the norm is
continuous across $\gamma = 1$. In contrast to the exceptional-point line, where
$\Vert A_\lambda\Vert^2 \sim L$, the $\mathcal{PT}$ transition is thus \emph{not}
marked by an extensive enhancement. Instead, the AGP norm saturates to a finite,
system-size- and field-independent constant. Physically this reflects the
vanishing of the transverse field at $g = 0$, where the state approaches a
classical Ising product configuration whose adiabatic preparation carries no
divergent counterdiabatic cost.

In the thermodynamic limit ($L\to\infty$ before $\gamma\to\gamma_c$) the two
Ising points and the $\mathcal{PT}$ point behave differently. Writing
$\nu=\ln(J/g)$ (unbroken) and $\nu_R=\ln(J/\mu)$ (broken), the large-$L$ asymptotics
of \eqref{eq:agp-norm-unbroken}--\eqref{eq:agp-norm-broken} are
\begin{align}
    \Vert A_\lambda\Vert^2 \;\xrightarrow{L\to\infty}\;
\begin{cases}
    &\frac{1}{64\,J^2\,\lvert\nu\rvert}\;\;(\text{Ising QCP, }g\to J) \\
    &\frac{1}{64\,J^2\,\nu_R}\;\;(\text{EP, }\mu\to J)
\end{cases},
\end{align}
where the $L$-dependent exponentials cancel identically. Since
$\lvert\nu\rvert\simeq\lvert g-J\rvert/J\propto\lvert\gamma-\gamma_c\rvert$ near
$g=J$, and likewise $\nu_R\propto\lvert\gamma-\gamma_c\rvert$ near $\mu=J$, both
Ising lines give a genuine power-law divergence
\begin{align}
    \Vert A_\lambda\Vert^2 \sim \lvert\gamma-\gamma_c\rvert^{-z},
    \qquad z = 1,
\end{align}
consistent with the Ising exponents $z=\nu_{\text{Ising}}=1$. At the $\mathcal{PT}$ transition,
by contrast, $g\to0$ drives $\nu\to\infty$ and the norm saturates to the finite
constant $1/(32J^2)$ of Eq.~\eqref{eq:agpnorm-scaling-PT}, so it neither diverges nor vanishes, giving $z=0$. The two Ising critical lines thus flank the $\mathcal{PT}$ point with divergent adiabatic susceptibility, while the $\mathcal{PT}$ point itself is regular, in contrast to the $\mathcal{PT}$ transition in the isotropic Heisenberg chain observed in Fig.~\ref{fig: AGP norm}. The NH TFIM has an all-bands-flat $\mathcal{PT}-$transition point (i.e., $E_k = 2 J$ at $\gamma = 1$), while the Heisenberg chain has a gap-closing transition.

\section{Conclusion}
In this work, we have developed a general, diagonalization-free framework for engineering shortcuts to adiabaticity in non-Hermitian systems by representing the adiabatic gauge potential in Krylov space. After introducing an integral representation of the counterdiabatic control, we expressed the AGP as a series of nested commutators with controlled locality and showed that the bi-Lanczos and Arnoldi algorithms provide an efficient basis in which its computation reduces to a sparse tridiagonal or upper-Hessenberg matrix equation. This construction generalizes the Hermitian Krylov-space formulation of Refs.~\cite{Takahashi_2024STA, Bhattacharjee_2023} to the biorthogonal setting, with the additional set of Lanczos coefficients $\{c_n\}$ reflecting the loss of Hermiticity. Truncating the series at finite order yields a systematically improvable approximate AGP whose accuracy can be benchmarked through the AGP norm.

We illustrated the versatility of the method across a range of physical systems. For a decaying two-level atom, the Krylov construction reproduces the exact counterdiabatic drive obtained by diagonalization and inherits the divergence of the control at the exceptional point. In the interacting Hatano-Nelson model, approximate controls built from only a small number of Krylov vectors strongly suppress the nonadiabatic excess energy and converge rapidly toward the exact result. For the $\mathcal{PT}-$symmetric Heisenberg chain, the AGP norm both converges quickly and serves as a sensitive probe of the $\mathcal{PT}-$symmetry-breaking transition. In every case, the expansion required only a small fraction of the full Krylov space dimension, underscoring the practical advantage of the approach for large systems where exact diagonalization is infeasible. We also discussed exact results for the AGP of the non-hermitian $\mathcal{PT}-$symmetric transverse field Ising model where we demonstrate that the AGP norm captures the Ising transitions in both the $\mathcal{PT}-$symmetric and $\mathcal{PT}-$broken phases.

Several directions remain open. A natural extension is to genuinely open dynamics governed by Lindbladian master equations, in which vectorization in a doubled Hilbert space would enable counterdiabatic control of mixed states within the same Krylov framework. The rapid convergence observed here also invites a closer study of how the AGP norm and its Krylov representation behave across integrable and chaotic regimes, where the scaling of the AGP with system size has been proposed as a diagnostic of quantum chaos. Finally, the controlled locality of the truncated controls makes the present scheme a promising candidate for implementation in monitored and postselected platforms that realize effective non-Hermitian dynamics, with potential applications to quantum optimization, state preparation, and sensing.

\acknowledgements
This project was supported by the Luxembourg National Research Fund (FNR Grant No 
C24/MS/18940482/STAOpen).

\section*{Data Availability}
Codes for numerical simulations are available in Github repository \cite{code}.

\appendix 
\section{Integral representation} \label{Integral_rep}
The adiabatic gauge potential has the following expression in the instantaneous biorthogonal eigenbasis of $H(\lambda)$ 
\begin{align}
    A_\lambda = -i\sum_n \ket{R_n}\bra{\partial_\lambda L_n}\,.
\end{align}
The matrix elements of this operator in the same basis is given by
\begin{align}
    \bra{L_m}A_\lambda \ket{R_n}\vert_{m\neq n} = -i\braket{\partial_\lambda L_m}{R_n}\,.
\end{align}
This can be in turn written as 
\begin{align}
    -i\braket{\partial_\lambda L_m}{R_n} = -i\frac{\bra{L_m}\partial_\lambda H \ket{R_n}}{E_m - E_n}\,,
\end{align}
where in the last step the Feynman-Hellmann theorem is used. Note that $E_m$ are in general complex. Let us now consider the eigen-decomposition of $H(\lambda)$. This is given by $H = \sum_{n}E_n(\lambda) \ket{R_n}\bra{L_n}$. In particular, we also require the Hermitian conjugate of this $H^\dagger = \sum_{n}E^{*}_n\ket{L_n}\bra{R_n}$. Let us denote the corresponding propagrators as $U_{-}(t) = \exp(-i H t)$ and $U_{+}(t) = \exp(i H t) = (\exp(-i H^\dagger t))^\dagger$. The time-evolution operator
\begin{align}
    A_\lambda =-\frac{1}{2}\int_{-\infty}^{\infty}\mathrm{d}s\,\text{sgn}(s)U_{+}(s) \partial_\lambda H U_{-}(s)\,,
\end{align}
which has the following matrix elements
\begin{align}
    \bra{L_m}A_\lambda \ket{R_n} &= -\frac{1}{2}\int_{-\infty}^{\infty}\mathrm{d}s\,\text{sgn}(s)e^{i (E_m-E_n) s}\\
    &\quad \times\bra{L_m}\partial_\lambda H_\lambda \ket{R_n}\notag\\ &=-i\frac{\bra{L_m}\partial_\lambda H_\lambda \ket{R_n}}{E_m - E_n}\,.
\end{align}
The integral expression can also be adjusted by regularization $\mu \ll 1$. The regularized integral has the matrix elements
\begin{align}
   & \bra{L_m} A_\lambda \ket{R_n} \nonumber \\
    &= -\frac{1}{2} \lim_{\mu \to 0} \int_{-\infty}^\infty {\rm sgn}(s)e^{-\mu |s|}\, \bra{L_m}e^{iH_\lambda s} \partial_\lambda He^{-iH_\lambda s}\ket{R_n}\, ds \nonumber\\
    & = -\frac{1}{2} \bra{L_m}\partial_\lambda H\ket{R_n}\lim_{\mu \to 0} \int_{-\infty}^\infty {\rm sgn}(s)e^{-\mu |s|}\, e^{-i(E_n - E_m) s} \, ds\,,
\end{align}
where we have used the eigenvalue relations in Eq. \eqref{time-indep schrodinger eqn}. In the limit $\mu \to 0$, the integral vanishes identically in the case of $m = n$. Else, we can define $\omega_{nm} = E_n - E_m$, and split the integral into two parts for $s<0$ and $s>0$
\begin{align}
  I &=  \lim_{\mu \to 0}\left[\int_0^\infty e^{-(\mu + i \omega_{nm})s}\;ds - \int_0^\infty e^{-(\mu - i \omega_{nm})s}\,ds \right]\nonumber \\
  & =  \lim_{\mu \to 0} \left[\frac{1}{\mu + i \omega_{nm}} - \frac{1}{\mu - i \omega_{nm}}\right] = \frac{2i}{E_m - E_n}\,.
\end{align}
Plugging this back gives the expected equation
\begin{align}
    \bra{L_m} A_\lambda \ket{R_n} = -i \frac{\bra{L_m}\partial_\lambda H\ket{R_n}}{(E_m - E_n)}\,.
\end{align}
Therefore, the integral representation of the adiabatic gauge potential involves a two-Hamiltonian dynamics, with the forward evolution generated by $H$ and the backward evolution generated by $H^\dagger$. If the regularization $\mu$ is retained in the expression of $I$, the resulting term is $2 i \omega_{m n}/(\omega_{m n}^2 + \mu^2)$ which handles any accidental degeneracies.

\section{Krylov basis and bi-Lanczos Algorithm }\label{sec:bilanczos}
Krylov basis arise naturally in the context of time-evolution of an operator $\mathcal{O}_0$. Consider the evolution of the operator in the Heisenberg picture
\begin{align}
    \mathcal{O}(t) = e^{i\mathcal{H}t} \mathcal{O}_0 e^{-i\mathcal{H}t} = \sum_{n = 0}^\infty \frac{(it)^n}{n!} \hat{\mathcal{L}}^n \mathcal{O}_0\,,
\end{align}
where $\hat{\mathcal{L}}^n\mathcal{O}_0 = [\mathcal{H}, [\mathcal{H}, \dots,[\mathcal{H}, \mathcal{O}_0]]]$ with $n$ nested commutators. A basis can be defined using the terms in the series
\begin{align}
    \mathcal{B} = \rm{span}\{\hat{\mathcal{L}}^n \mathcal{O}_0\}_{n=0}^\infty\,,
\end{align}
which is not naturally orthonormal. For the evolution by a Hermitian Hamiltonian $\mathcal{H}$, the well-known Lanczos algorithm can be used to iteratively orthonormalize the basis \cite{viswanath1994recursion}. Starting with the normalized initial operator $|\mathcal{O}_0)$, where the normalization is with respect to Frobenius inner product $(X|Y) = \frac{1}{\mathcal{D}}\rm{Tr}(X^\dagger Y)$, the basis vectors are generated iteratively using the algorithm 
\begin{align}
    b_n|\mathcal{O}_n) = \hat{\mathcal{L}}|\mathcal{O}_{n-1}) - b_{n-1}| \mathcal{O}_{n-2})\,,
\end{align}
where $b_n = \sqrt{(\mathcal{O}_n |\mathcal{O}_n)}$ and $b_{-1} = 0$. Here, $\mathcal{D}$ in the normalization factor is the dimension of the Hilbert space generated by the Hamiltonian. The algorithm halts when $b_n = 0$. The new basis $\mathcal{K} \equiv \{|\mathcal{O}_0), |\mathcal{O}_1), \dots,|\mathcal{O}_{K-1})\}$ is called Krylov basis, and for finite dimensional Hilbert space, its dimension satisfies the relation $K \leq \mathcal{D}^2 - \mathcal{D}+1$ \cite{Rabinovici2020opcomplexity}. The time-evolved operator $|\mathcal{O}(t))$ can be written in this basis as
\begin{align}
    |\mathcal{O}(t)) = \sum_{n = 0}^{K-1} i^n\varphi_n(t) |\mathcal{O}_n)\,,
\end{align}
and the superoperator $\hat{\mathcal{L}}$ takes the tridiagonal form 
\begin{align}
    (\mathcal{O}_m|\hat{\mathcal{L}}|\mathcal{O}_n) =
\begin{pmatrix}
0 & b_1 & 0   & 0    & \cdots \\
b_1 & 0 & b_2 & 0    & \cdots \\
0   & b_2 & 0 & b_3  & \cdots \\
0   & 0 & b_3 & 0  & \cdots \\
\vdots & \vdots & \vdots & \vdots & \ddots
\end{pmatrix}\,.
\end{align}
The Krylov basis provides a minimal set of basis vectors to study the time evolution of an operator $\mathcal{O}_0$. The ``Operator Growth Hypothesis" suggests that the set of Lanczos coefficients $\{b_n\}$, and Krylov Complexity
\begin{align}
    C(t) = \sum_{n = 0}^{K-1}n |\varphi_n(t)|^2\,, 
\end{align}
can be used to probe chaotic dynamics \cite{Parker2019opgrowth}. For chaotic models, the Lanczos coefficients show linear growth $b_n \sim \alpha n + \gamma$ (with some assumptions), which results in exponential growth of Krylov Complexity $C(t) \sim e^{2 \alpha t}$ at early times \cite{barbon2019evolution, Parker2019opgrowth, Rabinovici2020opcomplexity, Bhattacharjee2022saddle}. 

However, the simplicity of the algorithm relies on the Hermitian properties of the Hamiltonian $\mathcal{H}$. For non-Hermitian systems, the orthogonalization procedure requires a more general algorithm like Arnoldi iteration or the bi-Lanczos algorithm. These algorithms have been used extensively to study Krylov complexity in non-Hermitian contexts \cite{Bhattacharya2022arnoldi, Bhattacharjee2023dissSYK, 
Nizami2023Floquet,Nizami2024Floquet,Bhattacharya2023biLanczos,Zhou2025manybodychaos, baggioli2026nonhermitian, bhattacharyya2025kcopenquantum}. We will first focus on bi-Lanczos algorithm, which can be used to generate a minimal biorthonormal basis iteratively as follows. Starting with a general operator $|P_0)$ and its biorthogonal partner $(Q_0|$ (such that $(Q_0|P_0) = 1$), the biorthonormal basis can be generated with the iterative algorithm
\begin{align}
    &|P_{j+1}) = \frac{1}{c_{j+1}} [\mathcal{L} |P_j) - a_j |P_j) - b_j|P_{j-1})]\,, \nonumber \\ 
    &(Q_{j+1}| = \frac{1}{b_{j+1}} [(Q_j|\mathcal{L} - a_j (Q_j| - c_j(Q_{j-1}|]\,, \label{biLanczos}
\end{align}
where the coefficients $a_j$, $b_j$, and $c_j$ are analogous to the Lanczos coefficients. In detail,
\begin{enumerate}
    \item With the given initial operators, define $|R_0) = \mathcal{L}|P_0) - a_0 |P_0)$ and  $(S_0| = (Q_0|\mathcal{L} - a_0 (Q_0|$, where $a_0 = (Q_0|\mathcal{L}|P_0)$.

    \item For $j \in [1,2,\dots]$, run the following steps:
    \begin{enumerate}[label = (\alph*)]
        \item Calculate $\omega_j = (S_{j-1}|R_{j-1})$, $c_j = \sqrt{|\omega_j|}$, and $b_j = \omega_j/c_j$.

        \item  If $c_j \neq 0$, define new Krylov basis vectors
        \begin{align}
            |P_j) = \frac{|R_{j-1})}{c_j}\;\; \& \;\;(Q_j| = \frac{(S_{j-1}|}{b_j}\,.
        \end{align}

        \item Calculate the new intermediate vectors
        \begin{align}
            |R_j) &= \mathcal{L}|P_j) - a_j |P_j) - b_j |P_{j-1})\,, \nonumber\\  
            (S_j| &= (Q_j|\mathcal{L} - a_j(Q_j|  - c_j (Q_{j-1}|\,,
        \end{align}
        where, $a_j = (Q_j|\mathcal{L}|P_j)$ and go back to (a).
    \end{enumerate}
    \item Halt the algorithm if $c_j = 0$ and the resultant set of vectors $\{|P_0), |P_1), ..., |P_{K-1})\}$ and $\{(Q_0|, (Q_1|, ..., (Q_{K-1}|\}$ are Krylov basis vectors satisfying the bi-orthonormal condition.    
\end{enumerate}
In the new basis, the superoperator $\mathcal{L}$ takes the tridiagonal form 
\begin{align}
    (Q_m| \mathcal{L}|P_n) = \begin{pmatrix}
a_0 & b_1 & 0   & 0    & \cdots \\
c_1 & a_1 & b_2 & 0    & \cdots \\
0   & c_2 & a_2 & b_3  & \cdots \\
0   & 0 & c_3 & a_3  & \cdots \\
\vdots & \vdots & \vdots & \vdots & \ddots
\end{pmatrix}\,. 
\end{align}
The coefficients satisfy $c_j \in \mathbb{R}$  and $|b_j| = |c_j|$ by construction. Although the tridiagonal representation of $\mathcal{L}$ in the bi-orthogonal Krylov basis is computationally cost-effective, the algorithm is known to be numerically unstable. An alternative is to use the Arnoldi iteration, in which $\mathcal{L}$ takes an upper Hessenberg form and only requires one set of orthonormal vectors.

\section{Krylov basis using Arnoldi iteration}\label{sec:arnoldi}
Arnoldi iteration is a generalization of the Lanczos algorithm that can be used to generate an orthonormal Krylov basis for a non-Hermitian $\mathcal{L}$.  Starting with a normalized vector $|K_0)$, the orthogonal basis can be generated iteratively using the algorithm 
\begin{align}
    |A_n) &= \mathcal{L}|K_{n-1}) - \sum_{j=0}^{n-1}h_{j,n-1}|K_j)\,, \nonumber \\
    |K_n) &= \frac{1}{\sqrt{(A_n|A_n)}} |A_n) \equiv \frac{1}{h_{n,n-1}}|A_n)\,,\label{eq:arnoldi-alg}
\end{align}
where $h_{j,k} = (K_j|\mathcal{L}|K_k)$ are the Arnoldi coefficients. The algorithm halts naturally once the complete set of basis vectors $\{|K_n)\}$ are generated, and $h_{n,n-1} = 0$. For Lanczos and bi-Lanczos algorithms, each iterative step requires orthogonalization with only the two previous vectors. However, Arnoldi iteration requires orthogonalization with all previously generated vectors, making it costly but robust against numerical errors. Moreover, it requires less memory compared to bi-Lanczos since only one set of basis vectors is generated. In this basis, $\mathcal{L}$ takes the upper-Hessenberg form
\begin{align}
    (K_m|\mathcal{L}|K_n) = \begin{pmatrix}
h_{0,0} & h_{0,1} & h_{0,2}   & h_{0,3}    & \cdots \\
h_{1,0} & h_{1,1} & h_{1,2} & h_{1,3}    & \cdots \\
0   & h_{2,1} & h_{2,2} & h_{2,3}  & \cdots \\
0   & 0 & h_{3,2} & h_{3,3}  & \cdots \\
\vdots & \vdots & \vdots & \vdots & \ddots
\end{pmatrix}\,. 
\end{align}
with the same upper bound on the Krylov space dimension $K\leq \mathcal{D}^2-\mathcal{D}+1$. We will now use the Krylov basis to write an efficient representation of the AGP.

\section{Krylov basis for decaying two-level atoms} \label{Krylov basis}
Using the Pauli basis and Frobenius inner product, the explicit form of the superoperator $\mathcal{L}$ and $\partial_t H$ can be written as
\begin{align}
    L =
\begin{pmatrix}
0 & 0 & 0 & 0 \\
0 & 0 & \alpha & 0 \\
0 & -\alpha & 0 & -\beta \\
0 & 0 & \beta & 0
\end{pmatrix},\;\;\;\;
\ket{\partial_tH} = \begin{pmatrix}
    \nu \\\gamma \\0 \\ \delta
\end{pmatrix}\,,
\end{align}
with 
\begin{align}
    &\alpha = \Gamma + 2i\Delta,&\beta = 2i\Omega, &&  \nonumber\\
    &\nu = -{i\dot{\Gamma}}/2b_0, & \gamma = \dot{\Omega}/b_0, && \delta = -\dot{\Delta}/b_0 + {i\dot{\Gamma}}/2b_0 \,,\nonumber
\end{align}
 where $b_0 = \sqrt{\dot{\Omega}^2 + \dot{\Delta}^2 + \dot{\Gamma}^2/2}$ is the normalization factor. We start with $\ket{P_0} = \ket{Q_0} = \begin{pmatrix}
    \nu, & \gamma, & 0, &\delta
\end{pmatrix}^T$. The next set of Krylov basis vectors can be calculated using Eq. \eqref{biLanczos} as
\begin{align}
    \ket{P_1} = -\frac{1}{c_1}\begin{pmatrix}
        0 \\ 0 \\ \alpha \gamma + \beta \delta \\0
    \end{pmatrix},\;\;\;\; \ket{Q_1} = \frac{1}{b_1^*}\begin{pmatrix}
        0 \\ 0 \\ \alpha^* \gamma + \beta^* \delta \\0
    \end{pmatrix}\,,
\end{align}
where $\omega_1 = -(\alpha\gamma + \beta \delta)^2$, $c_1 = |\alpha\gamma + \beta \delta|$ and $b_1 = \omega_1/c_1$. The diagonal element $a_0 = 0$. We can iteratively calculate the third set of basis vectors as
\begin{align}
    \ket{P_2} = -\frac{1}{c_2}\begin{pmatrix}
        -b_1 \nu \\ -\alpha - b_1 \gamma \\ 0 \\-\beta - b_1 \delta
    \end{pmatrix},\;\;\;\; \ket{Q_2} = \frac{1}{b_2^*}\begin{pmatrix}
        -c_1 \nu \\ \alpha^* - c_1\gamma \\ 0 \\\beta^* - c_1 \delta
    \end{pmatrix}\,,
\end{align}
where $\omega_2 = -\alpha^2 - \beta^2 - \omega_1$. Lanczos coefficients $c_2$ and $b_2$ can be calculated as before, and it turns out that $a_1 = 0$. Upon continuing the iterative process, we find $a_2 = 0$, and $\ket{P_3} \propto \vec{0}$. The algorithm terminates in this step, resulting in a three dimensional Krylov space.

\section{Krylov basis for non-Hermitian transverse field Ising model} \label{Krylov nhtfim}
The details of calculations for the Krylov basis for the non-Hermitian transverse field Ising model (NH-TFIM) are presented. The basis vectors are strings of spins $\tau^{x,y,z}_i$ that map to fermion bilinears under Jordan-Wigner transformation~\cite{Takahashi_2024STA}. These are given by 
\begin{align}
    M &= \sum_{j = 1}^{L}\tau^{z}_j\,,\\
    V^{x}_k &= \sum_{j=1}^{L}\tau^{x}_jZ_{j,k}\tau^{x}_{j+k} \,,\\
    V^{y}_k &= \sum_{j=1}^{L}\tau^{y}_jZ_{j,k}\tau^{y}_{j+k}\,,\\
    W_{k} &= \frac{1}{\sqrt{2}} \sum_{j=1}^{L} \left(\tau^{x}_jZ_{j,k}\tau^{y}_{j+k} + \tau^{y}_jZ_{j,k}\tau^{x}_{j+k}\right)\,,
\end{align}
Each operator has the Hilbert-Schmidt norm $L$. We consider the normalized fermion bilinears by dividing each by $\sqrt{L}$. The action of the term Liouvillian on the fermion bilinears is given by
\begin{align}
    \mathcal{L}_g M &= 2\sqrt{2}i J W_1\,,\label{eq:LgM}\\
    \mathcal{L}_g V^{x}_k &= 2\sqrt{2}i \left( J W_{k-1} + g W_k\right)\,,\label{eq:LgVx}\\
    \mathcal{L}_g V^{y}_k &= 2\sqrt{2}i \left(-J W_{k+1} - g W_k\right)\,,\label{eq:LgVy}\\
    \mathcal{L}_g W_k &= 2\sqrt{2} i J [ V^{y}_{k-1} - V^{x}_{k+1} - \frac{g}{J}(V^{x}_k - V^{y}_k) - \delta_{k,1}M]\,.\label{eq:LgWk}
\end{align}
Following the algorithm in Appendix~\ref{sec:bilanczos}, we can start with the left and right initial vectors $\vert P_0 ) = \vert Q_0 ) = M/\sqrt{L} \equiv \hat{O}$. This gives us $a_0 = (Q_0\vert \mathcal{L}_g \vert P_0) = 0$. The intermediate vectors are $|S_0) = |R_0) = 2\sqrt{2} i J W_1/\sqrt{L}$, using Eq.~\eqref{eq:LgM}. This gives us the first bi-Lanczos coefficient as
\begin{align}
    \omega_1 &= (S_0\vert R_0) = 8 J^2,\,c_1 = b_1 = 2\sqrt{2}J,\notag\\ \,\vert P_1) &= \vert Q_1) = i W_1/\sqrt{L}\,.
\end{align}
The next step of the iteration can be computed by using Eq.~\eqref{eq:LgWk} which gives us
\begin{align}
    \vert R_1) = 2\sqrt{2}(J V^{x}_2 + i\mu V^{x}_1 - i\mu V^{y}_1)/\sqrt{L}\,,\\
    \vert S_1) = 2\sqrt{2}(J V^{x}_2 - i\mu V^{x}_1 + i\mu V^{y}_1)/\sqrt{L}\,.
\end{align}
The corresponding coefficient is $\omega_2 = (S_1 \vert R_1) = 8(J^2 - 2 \mu^2)$. The Krylov vectors can then be computed to give
\begin{align}
    \vert P_2 ) =& \frac{i\mu V^{x}_1 + J V^{x}_2 - i\mu V^{y}_1}{\sqrt{L\vert J^2 - 2\mu^2\vert }}\,,\\
    \vert Q_2 ) =& \frac{-i\mu V^{x}_1 + J V^{x}_2 + i\mu V^{y}_1}{\text{sgn}(J^2 - 2\mu^2)\sqrt{L\vert J^2 - 2\mu^2\vert }}\,.
\end{align}
The third step can be computed using Eq.~\eqref{eq:LgVx}-\eqref{eq:LgVy}. The resulting vectors are 
\begin{align}
    \vert R_2 ) &= -\frac{4\sqrt{2}J\mu}{\sqrt{L\vert J^2 - 2\mu^2\vert}}W_2\,,\\
    \vert S_2 ) &= \frac{4\sqrt{2} J\mu}{\text{sgn}(J^2 - 2\mu^2)\sqrt{L\vert J^2 - 2\mu^2\vert }}W_2\,.
\end{align}
This leads us to the basis vectors $\vert P_3 ) = \vert Q_3 ) = - W_2 /\sqrt{L}$ and $\omega_3 = (-32 J^2 \mu^2)/(J^2 - 2\mu^2)$. Following this, it is reasonable to assume that $\vert P_{2k-1}) = \vert Q_{2k-1}) = i^{k} W_{k}/\sqrt{L}$. Let us verify this by the application of the generator $\mathcal{L}_g$ twice. Using the bi-Lanczos algorithm, we find that
\begin{eqnarray}
    \mathcal{L}^2_g \vert P_{2k-1}) &=& c_{2k}c_{2k+1}\vert P_{2k+1})+ b_{2k-1}b_{2k-2}\vert P_{2k-3}) \notag\\& &+ (c_{2k-1}b_{2k-1} + c_{2k}b_{2k})\vert P_{2k-1})\,.
\end{eqnarray}
Using Eqs.~\eqref{eq:LgM}-\eqref{eq:LgWk}, we obtain the constraint relations
\begin{align}
    b_{2k-1}c_{2k-1} + b_{2k}c_{2k} &= 16 J^2 \left(1+ \frac{g^2}{J^2}\right)\,,\\
    b_{2k}b_{2k + 1} = c_{2k+2}c_{2k+3} &= -16 J g\,.
\end{align}
These constraint relations are the generalisation of Eqs.~\eqref{eq:bn-const-1}-\eqref{eq:bn-const-2} to the bi-Lanczos basis. These are valid in both $\mathcal{PT}-$ broken and unbroken phases. The resulting matrix on the LHS of Eq.~\eqref{Mat_eq_biLanczos} is therefore a Toeplitz matrix.

\vfill

\bibliography{STA_Krylov}

@article{Alipour2022,
  title = {Entropy-based formulation of thermodynamics in arbitrary quantum evolution},
  author = {Alipour, S. and Rezakhani, A. T. and Chenu, A. and del Campo, A. and Ala-Nissila, T.},
  journal = {Phys. Rev. A},
  volume = {105},
  issue = {4},
  pages = {L040201},
  numpages = {6},
  year = {2022},
  month = {Apr},
  publisher = {American Physical Society},
  doi = {10.1103/PhysRevA.105.L040201},
  url = {https://link.aps.org/doi/10.1103/PhysRevA.105.L040201}
}

@article{Dupays2021,
  doi = {10.22331/q-2021-05-01-449},
  url = {https://doi.org/10.22331/q-2021-05-01-449},
  title = {Shortcuts to {S}queezed {T}hermal {S}tates},
  author = {Dupays, L{\'{e}}once and Chenu, Aur{\'{e}}lia},
  journal = {{Quantum}},
  issn = {2521-327X},
  publisher = {{Verein zur F{\"{o}}rderung des Open Access Publizierens in den Quantenwissenschaften}},
  volume = {5},
  pages = {449},
  month = may,
  year = {2021}
}

@article{Tanaka2012,
  title = {Robust adaptive measurement scheme for qubit-state preparation},
  author = {Tanaka, Saki and Yamamoto, Naoki},
  journal = {Phys. Rev. A},
  volume = {86},
  issue = {6},
  pages = {062331},
  numpages = {6},
  year = {2012},
  month = {Dec},
  publisher = {American Physical Society},
  doi = {10.1103/PhysRevA.86.062331},
  url = {https://link.aps.org/doi/10.1103/PhysRevA.86.062331}
}

@article{Hegade2022,
  title = {Digitized counterdiabatic quantum optimization},
  author = {Hegade, Narendra N. and Chen, Xi and Solano, Enrique},
  journal = {Phys. Rev. Res.},
  volume = {4},
  issue = {4},
  pages = {L042030},
  numpages = {7},
  year = {2022},
  month = {Nov},
  publisher = {American Physical Society},
  doi = {10.1103/PhysRevResearch.4.L042030},
  url = {https://link.aps.org/doi/10.1103/PhysRevResearch.4.L042030}
}

@article{Chandarana2022,
  title = {Digitized-counterdiabatic quantum approximate optimization algorithm},
  author = {Chandarana, P. and Hegade, N. N. and Paul, K. and Albarr\'an-Arriagada, F. and Solano, E. and del Campo, A. and Chen, Xi},
  journal = {Phys. Rev. Res.},
  volume = {4},
  issue = {1},
  pages = {013141},
  numpages = {9},
  year = {2022},
  month = {Feb},
  publisher = {American Physical Society},
  doi = {10.1103/PhysRevResearch.4.013141},
  url = {https://link.aps.org/doi/10.1103/PhysRevResearch.4.013141}
}

@article{Dann2019,
  title = {Shortcut to Equilibration of an Open Quantum System},
  author = {Dann, Roie and Tobalina, Ander and Kosloff, Ronnie},
  journal = {Phys. Rev. Lett.},
  volume = {122},
  issue = {25},
  pages = {250402},
  numpages = {6},
  year = {2019},
  month = {Jun},
  publisher = {American Physical Society},
  doi = {10.1103/PhysRevLett.122.250402},
  url = {https://link.aps.org/doi/10.1103/PhysRevLett.122.250402}
}

@Article{McArdle2019,
author={McArdle, Sam
and Jones, Tyson
and Endo, Suguru
and Li, Ying
and Benjamin, Simon C.
and Yuan, Xiao},
title={Variational ansatz-based quantum simulation of imaginary time evolution},
journal={npj Quantum Information},
year={2019},
month={Sep},
day={06},
volume={5},
number={1},
pages={75},
issn={2056-6387},
doi={10.1038/s41534-019-0187-2},
url={https://doi.org/10.1038/s41534-019-0187-2}
}

@Article{Motta2020,
author={Motta, Mario
and Sun, Chong
and Tan, Adrian T. K.
and O'Rourke, Matthew J.
and Ye, Erika
and Minnich, Austin J.
and Brand{\~a}o, Fernando G. S. L.
and Chan, Garnet Kin-Lic},
title={Determining eigenstates and thermal states on a quantum computer using quantum imaginary time evolution},
journal={Nature Physics},
year={2020},
month={Feb},
day={01},
volume={16},
number={2},
pages={205-210},
issn={1745-2481},
doi={10.1038/s41567-019-0704-4},
url={https://doi.org/10.1038/s41567-019-0704-4}
}

@article{Ding2023,
  title = {Fundamental Sensitivity Limits for Non-Hermitian Quantum Sensors},
  author = {Ding, Wenkui and Wang, Xiaoguang and Chen, Shu},
  journal = {Phys. Rev. Lett.},
  volume = {131},
  issue = {16},
  pages = {160801},
  numpages = {7},
  year = {2023},
  month = {Oct},
  publisher = {American Physical Society},
  doi = {10.1103/PhysRevLett.131.160801},
  url = {https://link.aps.org/doi/10.1103/PhysRevLett.131.160801}
}

@Article{Yin2022,
author={Yin, Zelong
and Li, Chunzhen
and Allcock, Jonathan
and Zheng, Yicong
and Gu, Xiu
and Dai, Maochun
and Zhang, Shengyu
and An, Shuoming},
title={Shortcuts to adiabaticity for open systems in circuit quantum electrodynamics},
journal={Nature Communications},
year={2022},
month={Jan},
day={10},
volume={13},
number={1},
pages={188},
issn={2041-1723},
doi={10.1038/s41467-021-27900-6},
url={https://doi.org/10.1038/s41467-021-27900-6}
}

@Article{Lau2018,
author={Lau, Hoi-Kwan
and Clerk, Aashish A.},
title={Fundamental limits and non-reciprocal approaches in non-Hermitian quantum sensing},
journal={Nature Communications},
year={2018},
month={Oct},
day={17},
volume={9},
number={1},
pages={4320},
issn={2041-1723},
doi={10.1038/s41467-018-06477-7},
url={https://doi.org/10.1038/s41467-018-06477-7}
}

@article{Morawetz2025,
  title = {Universal Counterdiabatic Driving in {K}rylov Space},
  author = {Morawetz, Stewart and Polkovnikov, Anatoli},
  journal = {PRX Quantum},
  volume = {6},
  issue = {4},
  pages = {040320},
  numpages = {13},
  year = {2025},
  month = {Oct},
  publisher = {American Physical Society},
  doi = {10.1103/wbbs-s8fs},
  url = {https://link.aps.org/doi/10.1103/wbbs-s8fs}
}

@article{Takahashi2013,
  title = {Transitionless quantum driving for spin systems},
  author = {Takahashi, Kazutaka},
  journal = {Phys. Rev. E},
  volume = {87},
  issue = {6},
  pages = {062117},
  numpages = {9},
  year = {2013},
  month = {Jun},
  publisher = {American Physical Society},
  doi = {10.1103/PhysRevE.87.062117},
  url = {https://link.aps.org/doi/10.1103/PhysRevE.87.062117}
}

@article{Takahashi2015,
  title = {Unitary deformations of counterdiabatic driving},
  author = {Takahashi, Kazutaka},
  journal = {Phys. Rev. A},
  volume = {91},
  issue = {4},
  pages = {042115},
  numpages = {8},
  year = {2015},
  month = {Apr},
  publisher = {American Physical Society},
  doi = {10.1103/PhysRevA.91.042115},
  url = {https://link.aps.org/doi/10.1103/PhysRevA.91.042115}
}

@Article{McDonald2020,
author={McDonald, Alexander
and Clerk, Aashish A.},
title={Exponentially-enhanced quantum sensing with non-Hermitian lattice dynamics},
journal={Nature Communications},
year={2020},
month={Oct},
day={23},
volume={11},
number={1},
pages={5382},
issn={2041-1723},
doi={10.1038/s41467-020-19090-4},
url={https://doi.org/10.1038/s41467-020-19090-4}
}

@article{Chandarana2023,
  title = {Digitized Counterdiabatic Quantum Algorithm for Protein Folding},
  author = {Chandarana, Pranav and Hegade, Narendra N. and Montalban, Iraitz and Solano, Enrique and Chen, Xi},
  journal = {Phys. Rev. Appl.},
  volume = {20},
  issue = {1},
  pages = {014024},
  numpages = {16},
  year = {2023},
  month = {Jul},
  publisher = {American Physical Society},
  doi = {10.1103/PhysRevApplied.20.014024},
  url = {https://link.aps.org/doi/10.1103/PhysRevApplied.20.014024}
}

@article{Hegade2021,
  title = {Shortcuts to Adiabaticity in Digitized Adiabatic Quantum Computing},
  author = {Hegade, Narendra N. and Paul, Koushik and Ding, Yongcheng and Sanz, Mikel and Albarr\'an-Arriagada, F. and Solano, Enrique and Chen, Xi},
  journal = {Phys. Rev. Appl.},
  volume = {15},
  issue = {2},
  pages = {024038},
  numpages = {13},
  year = {2021},
  month = {Feb},
  publisher = {American Physical Society},
  doi = {10.1103/PhysRevApplied.15.024038},
  url = {https://link.aps.org/doi/10.1103/PhysRevApplied.15.024038}
}

@Article{Chandarana2024,
author={Chandarana, Pranav
and Paul, Koushik
and Garcia-de-Andoin, Mikel
and Ban, Yue
and Sanz, Mikel
and Chen, Xi},
title={Photonic counterdiabatic quantum optimization algorithm},
journal={Communications Physics},
year={2024},
month={Sep},
day={30},
volume={7},
number={1},
pages={315},
issn={2399-3650},
doi={10.1038/s42005-024-01807-2},
url={https://doi.org/10.1038/s42005-024-01807-2}
}

@article{Torosov2013,
  title = {Non-Hermitian shortcut to adiabaticity},
  author = {Torosov, Boyan T. and Della Valle, Giuseppe and Longhi, Stefano},
  journal = {Phys. Rev. A},
  volume = {87},
  issue = {5},
  pages = {052502},
  numpages = {5},
  year = {2013},
  month = {May},
  publisher = {American Physical Society},
  doi = {10.1103/PhysRevA.87.052502},
  url = {https://link.aps.org/doi/10.1103/PhysRevA.87.052502}
}

@article{Ibanez2011,
  title = {Shortcuts to adiabaticity for non-Hermitian systems},
  author = {Ib\'a\~nez, S. and Mart\'{\i}nez-Garaot, S. and Chen, Xi and Torrontegui, E. and Muga, J. G.},
  journal = {Phys. Rev. A},
  volume = {84},
  issue = {2},
  pages = {023415},
  numpages = {8},
  year = {2011},
  month = {Aug},
  publisher = {American Physical Society},
  doi = {10.1103/PhysRevA.84.023415},
  url = {https://link.aps.org/doi/10.1103/PhysRevA.84.023415}
}

@article{Hornedal2025,
  doi = {10.22331/q-2025-05-05-1729},
  url = {https://doi.org/10.22331/q-2025-05-05-1729},
  title = {A geometrical description of non-{H}ermitian dynamics: speed limits in finite rank density operators},
  author = {H{\"{o}}rnedal, Niklas and Pro{\'{s}}niak, Oskar A. and del Campo, Adolfo and Chenu, Aur{\'{e}}lia},
  journal = {{Quantum}},
  issn = {2521-327X},
  publisher = {{Verein zur F{\"{o}}rderung des Open Access Publizierens in den Quantenwissenschaften}},
  volume = {9},
  pages = {1729},
  month = may,
  year = {2025}
}

@article{Erdamar2026,
  title = {Exploring the Riemann-Surface Topology of a Non-Hermitian Superconducting Qubit Using Shortcuts to Adiabaticity},
  author = {Erdamar, Serra and Abbasi, Maryam and Chen, Weijian and H\"ornedal, Niklas and Chenu, Aur\'elia and Murch, Kater W.},
  journal = {PRX Quantum},
  volume = {7},
  issue = {1},
  pages = {010337},
  numpages = {14},
  year = {2026},
  month = {Feb},
  publisher = {American Physical Society},
  doi = {10.1103/7gtf-4tbh},
  url = {https://link.aps.org/doi/10.1103/7gtf-4tbh}
}

@article{Dupays2025,
  title = {Slow approach to adiabaticity in many-body non-Hermitian systems: The Hatano-Nelson model},
  author = {Dupays, L\'eonce and del Campo, Adolfo and D\'ora, Bal\'azs},
  journal = {Phys. Rev. B},
  volume = {111},
  issue = {4},
  pages = {045130},
  numpages = {7},
  year = {2025},
  month = {Jan},
  publisher = {American Physical Society},
  doi = {10.1103/PhysRevB.111.045130},
  url = {https://link.aps.org/doi/10.1103/PhysRevB.111.045130}
}

@article{Dupays2020,
  title = {Superadiabatic thermalization of a quantum oscillator by engineered dephasing},
  author = {Dupays, L. and Egusquiza, I. L. and del Campo, A. and Chenu, A.},
  journal = {Phys. Rev. Res.},
  volume = {2},
  issue = {3},
  pages = {033178},
  numpages = {9},
  year = {2020},
  month = {Aug},
  publisher = {American Physical Society},
  doi = {10.1103/PhysRevResearch.2.033178},
  url = {https://link.aps.org/doi/10.1103/PhysRevResearch.2.033178}
}

@article{Chen2010,
  title = {Fast Optimal Frictionless Atom Cooling in Harmonic Traps: Shortcut to Adiabaticity},
  author = {Chen, Xi and Ruschhaupt, A. and Schmidt, S. and del Campo, A. and Gu\'ery-Odelin, D. and Muga, J. G.},
  journal = {Phys. Rev. Lett.},
  volume = {104},
  issue = {6},
  pages = {063002},
  numpages = {4},
  year = {2010},
  month = {Feb},
  publisher = {American Physical Society},
  doi = {10.1103/PhysRevLett.104.063002},
  url = {https://link.aps.org/doi/10.1103/PhysRevLett.104.063002}
}

@inbook{Torrontegui2013,
   title={Shortcuts to Adiabaticity},
   ISSN={1049-250X},
   url={http://dx.doi.org/10.1016/B978-0-12-408090-4.00002-5},
   DOI={10.1016/b978-0-12-408090-4.00002-5},
   booktitle={Advances in Atomic, Molecular, and Optical Physics},
   publisher={Elsevier},
   author={Torrontegui, Erik and Ibáñez, Sara and Martínez-Garaot, Sofia and Modugno, Michele and del Campo, Adolfo and Guéry-Odelin, David and Ruschhaupt, Andreas and Chen, Xi and Muga, Juan Gonzalo},
   year={2013},
   pages={117–169} }

@article{delCampo2012,
    title = {{Assisted Finite-Rate Adiabatic Passage Across a Quantum Critical Point: Exact Solution for the Quantum Ising Model}},
    author = {del Campo, Adolfo and Rams, Marek M. and Zurek, Wojciech H.},
    year = 2012,
    journal = {Phys. Rev. Lett.},
    publisher = {American Physical Society},
    volume = 109,
    pages = 115703,
    doi = {10.1103/PhysRevLett.109.115703},
    issue = 11,
    numpages = 5
}

@article{Saberi2014,
    title = {{Adiabatic tracking of quantum many-body dynamics}},
    author = {Saberi, Hamed and Opatrn\'y, Tom\'a\v{s} and M\o{}lmer, Klaus and del Campo, Adolfo},
    year = 2014,
    journal = {Phys. Rev. A},
    publisher = {American Physical Society},
    volume = 90,
    pages = {060301},
    doi = {10.1103/PhysRevA.90.060301},
    issue = 6,
    numpages = 5
}

@article{Damski2014,
    title = {{Counterdiabatic driving of the quantum Ising model}},
    author = {Bogdan Damski},
    year = 2014,
    journal = {J. Stat. Mech. Theory Exp.},
    publisher = {IOP Publishing and SISSA},
    volume = 2014,
    number = 12,
    pages = {P12019},
    doi = {10.1088/1742-5468/2014/12/P12019}
}

@article{Sels2017,
    title = {{Minimizing irreversible losses in quantum systems by local counterdiabatic driving}},
    author = {Dries Sels and Anatoli Polkovnikov},
    year = 2017,
    journal = {Proc. Natl. Acad. Sci. USA},
    volume = 114,
    number = 20,
    pages = {E3909-E3916},
    doi = {10.1073/pnas.1619826114}
}

@article{GueryOdelin2019,
	title = {Shortcuts to adiabaticity: Concepts, methods, and applications},
	author = {Gu\'ery-Odelin, D. and Ruschhaupt, A. and Kiely, A. and Torrontegui, E. and Mart\'{\i}nez-Garaot, S. and Muga, J. G.},
	journal = {Rev. Mod. Phys.},
	volume = {91},
	issue = {4},
	pages = {045001},
	numpages = {54},
	year = {2019},
	publisher = {American Physical Society},
	doi = {10.1103/RevModPhys.91.045001},
}

@article{Hatomura2024,
doi = {10.1088/1361-6455/ad38f1},
url = {https://doi.org/10.1088/1361-6455/ad38f1},
year = {2024},
month = {apr},
publisher = {IOP Publishing},
volume = {57},
number = {10},
pages = {102001},
author = {Hatomura, Takuya},
title = {Shortcuts to adiabaticity: theoretical framework, relations between different methods, and versatile approximations},
journal = {Journal of Physics B: Atomic, Molecular and Optical Physics},
}

@article{Berry_2009,
doi = {10.1088/1751-8113/42/36/365303},
url = {https://doi.org/10.1088/1751-8113/42/36/365303},
year = {2009},
month = {aug},
publisher = {},
volume = {42},
number = {36},
pages = {365303},
author = {Berry, M V},
title = {Transitionless quantum driving},
journal = {Journal of Physics A: Mathematical and Theoretical},
}

@article{Demirplak&Rice_2003,
author = {Demirplak, Mustafa and Rice, Stuart A.},
title = {Adiabatic Population Transfer with Control Fields},
journal = {The Journal of Physical Chemistry A},
volume = {107},
number = {46},
pages = {9937-9945},
year = {2003},
doi = {10.1021/jp030708a},
URL = { https://doi.org/10.1021/jp030708a},
}

@article{Demirplak&Rice_2005,
author = {Demirplak, Mustafa and Rice, Stuart A.},
title = {Assisted Adiabatic Passage Revisited},
journal = {The Journal of Physical Chemistry B},
volume = {109},
number = {14},
pages = {6838-6844},
year = {2005},
doi = {10.1021/jp040647w},
URL = {https://doi.org/10.1021/jp040647w},
}

@article{Demirplak&Rice_2008,
    author = {Demirplak, Mustafa and Rice, Stuart A.},
    title = {On the consistency, extremal, and global properties of counterdiabatic fields},
    journal = {The Journal of Chemical Physics},
    volume = {129},
    number = {15},
    pages = {154111},
    year = {2008},
    month = {10},
    issn = {0021-9606},
    doi = {10.1063/1.2992152},
    url = {https://doi.org/10.1063/1.2992152},
}

@phdthesis{pandey2021studies,
  title={Studies of non-equilibrium behavior of quantum many-body systems using the adiabatic eigenstate deformations},
  author={Pandey, Mohit},
  year={2021},
  school={Boston University},
url = {https://open.bu.edu/items/175ae7c3-2123-45c9-8005-5b736c016205}
}

@article{Feynman_1939,
  title = {Forces in Molecules},
  author = {Feynman, R. P.},
  journal = {Phys. Rev.},
  volume = {56},
  issue = {4},
  pages = {340--343},
  numpages = {0},
  year = {1939},
  month = {Aug},
  publisher = {American Physical Society},
  doi = {10.1103/PhysRev.56.340},
  url = {https://link.aps.org/doi/10.1103/PhysRev.56.340}
}

@article{Kolodrubetz_2017,
title = {Geometry and non-adiabatic response in quantum and classical systems},
journal = {Physics Reports},
volume = {697},
pages = {1-87},
year = {2017},
issn = {0370-1573},
doi = {https://doi.org/10.1016/j.physrep.2017.07.001},
author = {Michael Kolodrubetz and Dries Sels and Pankaj Mehta and Anatoli Polkovnikov},
}

@article{Takahashi_2024STA,
  title = {Shortcuts to Adiabaticity in {K}rylov Space},
  author = {Takahashi, Kazutaka and del Campo, Adolfo},
  journal = {Phys. Rev. X},
  volume = {14},
  issue = {1},
  pages = {011032},
  numpages = {23},
  year = {2024},
  month = {Feb},
  publisher = {American Physical Society},
  doi = {10.1103/PhysRevX.14.011032},
  url ={https://link.aps.org/doi/10.1103/PhysRevX.14.011032}
}

@article{Claeys_2019,
  title = {Floquet-Engineering Counterdiabatic Protocols in Quantum Many-Body Systems},
  author = {Claeys, Pieter W. and Pandey, Mohit and Sels, Dries and Polkovnikov, Anatoli},
  journal = {Phys. Rev. Lett.},
  volume = {123},
  issue = {9},
  pages = {090602},
  numpages = {7},
  year = {2019},
  month = {Aug},
  publisher = {American Physical Society},
  doi = {10.1103/PhysRevLett.123.090602},
  url = {https://link.aps.org/doi/10.1103/PhysRevLett.123.090602}
}

@article{Bhattacharjee_2023,
    author = "Bhattacharjee, Budhaditya",
    title = "{A Lanczos approach to the Adiabatic Gauge Potential}",
    eprint = "2302.07228",
    archivePrefix = "arXiv",
    month = "2",
    year = "2023",
journal = ""
}

@Article{Visuri2026,
author={Visuri, Anne-Maria
and Gomez Cadavid, Alejandro
and Bhargava, Balaganchi A.
and Romero, Sebasti{\'a}n V.
and Grabarits, Andr{\'a}s
and Chandarana, Pranav
and Solano, Enrique
and del Campo, Adolfo
and Hegade, Narendra N.},
title={Digitized counterdiabatic quantum critical dynamics},
journal={npj Quantum Information},
year={2026},
month={Mar},
day={13},
volume={12},
number={1},
pages={47},
issn={2056-6387},
doi={10.1038/s41534-026-01208-z},
url={https://doi.org/10.1038/s41534-026-01208-z}
}

@misc{Grabarits2025,
      title={Universal Defect Statistics in Counterdiabatic Quantum Critical Dynamics}, 
      author={András Grabarits and Adolfo del Campo},
      year={2025},
      eprint={2503.22212},
      archivePrefix={arXiv},
      primaryClass={quant-ph},
      url={https://arxiv.org/abs/2503.22212}, 
}

@article{Vacanti2014,
doi = {10.1088/1367-2630/16/5/053017},
url = {https://doi.org/10.1088/1367-2630/16/5/053017},
year = {2014},
month = {may},
publisher = {IOP Publishing},
volume = {16},
number = {5},
pages = {053017},
author = {Vacanti, G and Fazio, R and Montangero, S and Palma, G M and Paternostro, M and Vedral, V},
title = {Transitionless quantum driving in open quantum systems},
journal = {New Journal of Physics},
}

@article{HacohenGourgy18,
  title = {Incoherent Qubit Control Using the Quantum Zeno Effect},
  author = {Hacohen-Gourgy, S. and Garc\'{\i}a-Pintos, L. P. and Martin, L. S. and Dressel, J. and Siddiqi, I.},
  journal = {Phys. Rev. Lett.},
  volume = {120},
  issue = {2},
  pages = {020505},
  numpages = {6},
  year = {2018},
  month = {Jan},
  publisher = {American Physical Society},
  doi = {10.1103/PhysRevLett.120.020505},
  url = {https://link.aps.org/doi/10.1103/PhysRevLett.120.020505}
}

@article{Lewalle2024,
  title = {Optimal Zeno Dragging for Quantum Control: A Shortcut to Zeno with Action-Based Scheduling Optimization},
  author = {Lewalle, Philippe and Zhang, Yipei and Whaley, K. Birgitta},
  journal = {PRX Quantum},
  volume = {5},
  issue = {2},
  pages = {020366},
  numpages = {30},
  year = {2024},
  month = {Jun},
  publisher = {American Physical Society},
  doi = {10.1103/PRXQuantum.5.020366},
  url = {https://link.aps.org/doi/10.1103/PRXQuantum.5.020366}
}

@misc{delcampo2026,
      title={Shortcuts to Adiabaticity via Adaptive Quantum {Z}eno Measurements}, 
      author={Adolfo del Campo},
      year={2026},
      eprint={2602.17786},
      archivePrefix={arXiv},
      primaryClass={quant-ph},
      url={https://arxiv.org/abs/2602.17786}, 
}

@article{Muga2004,
title = {Complex absorbing potentials},
journal = {Physics Reports},
volume = {395},
number = {6},
pages = {357-426},
year = {2004},
issn = {0370-1573},
doi = {https://doi.org/10.1016/j.physrep.2004.03.002},
url = {https://www.sciencedirect.com/science/article/pii/S0370157304001218},
author = {J.G. Muga and J.P. Palao and B. Navarro and I.L. Egusquiza},
}

@article{Ibanez2012Erratum,
  title = {Erratum: Shortcuts to adiabaticity for non-{H}ermitian systems [{P}hys. {R}ev. {A} 84, 023415 (2011)]},
  author = {Ib\'a\~nez, S. and Mart\'{\i}nez-Garaot, S. and Chen, Xi and Torrontegui, E. and Muga, J. G.},
  journal = {Phys. Rev. A},
  volume = {86},
  issue = {1},
  pages = {019901},
  numpages = {1},
  year = {2012},
  month = {Jul},
  publisher = {American Physical Society},
  doi = {10.1103/PhysRevA.86.019901},
  url = {https://link.aps.org/doi/10.1103/PhysRevA.86.019901}
}

@article{Zhang_2019,
  title = {Non-{H}ermitian quantum systems and their geometric phases},
  author = {Zhang, Qi and Wu, Biao},
  journal = {Phys. Rev. A},
  volume = {99},
  issue = {3},
  pages = {032121},
  numpages = {7},
  year = {2019},
  month = {Mar},
  publisher = {American Physical Society},
  doi = {10.1103/PhysRevA.99.032121},
  url = {https://link.aps.org/doi/10.1103/PhysRevA.99.032121}
}

@article{Huang2025adiabatic,
    author = "Huang, Minyi and Lee, Ray-Kuang",
    title = "{Adiabatic theorem for non-Hermitian quantum systems with non-degenerate real eigenvalues: A proof following Kato's approach}",
    eprint = "2511.00968",
    archivePrefix = "arXiv",
    month = "11",
    year = "2025",
    journal = ""
}

@article{Hajong2024HFtheorem,
  title = {Hellmann-{F}eynman theorem in non-{H}ermitian systems},
  author = {Hajong, Gaurav and Modak, Ranjan and Mandal, Bhabani Prasad},
  journal = {Phys. Rev. A},
  volume = {109},
  issue = {2},
  pages = {022227},
  numpages = {13},
  year = {2024},
  month = {Feb},
  publisher = {American Physical Society},
  doi = {10.1103/PhysRevA.109.022227},
  url = {https://link.aps.org/doi/10.1103/PhysRevA.109.022227}
}

@book{viswanath1994recursion,
  title={The recursion method: application to many-body dynamics},
  author={Viswanath, VS and M{\"u}ller, Gerhard},
  year={1994},
  publisher={Springer},
  url = {https://books.google.co.in/books?id=X2Ug4w17rnMC}
}

@article{Rabinovici2020opcomplexity,
    author = "Rabinovici, E. and S{\'a}nchez-Garrido, A. and Shir, R. and Sonner, J.",
    title = "{Operator complexity: a journey to the edge of Krylov space}",
    primaryClass = "hep-th",
    doi = "10.1007/JHEP06(2021)062",
    journal = "JHEP",
    volume = "06",
    pages = "062",
    year = "2021",
}

@article{Parker2019opgrowth,
  title = {A Universal Operator Growth Hypothesis},
  author = {Parker, Daniel E. and Cao, Xiangyu and Avdoshkin, Alexander and Scaffidi, Thomas and Altman, Ehud},
  journal = {Phys. Rev. X},
  volume = {9},
  issue = {4},
  pages = {041017},
  numpages = {29},
  year = {2019},
  month = {Oct},
  publisher = {American Physical Society},
  doi = {10.1103/PhysRevX.9.041017},
  url = {https://link.aps.org/doi/10.1103/PhysRevX.9.041017}
}

@article{Pratik2025Krylovreview,
title = {Quantum dynamics in {K}rylov space: {M}ethods and applications},
journal = {Physics Reports},
volume = {1125-1128},
pages = {1-82},
year = {2025},
issn = {0370-1573},
doi = {https://doi.org/10.1016/j.physrep.2025.05.001},
url = {https://www.sciencedirect.com/science/article/pii/S0370157325001462},
author = {Pratik Nandy and Apollonas S. Matsoukas-Roubeas and Pablo Martínez-Azcona and Anatoly Dymarsky and Adolfo {del Campo}},
}

@article{rabinovici2025krylovcomplexity,
      title={Krylov Complexity}, 
      author={Eliezer Rabinovici and Adrián Sánchez-Garrido and Ruth Shir and Julian Sonner},
      year={2025},
      eprint={2507.06286},
      archivePrefix={arXiv},
      url={https://arxiv.org/abs/2507.06286}, 
      journal = {}
}

@article{barbon2019evolution,
  title={On the evolution of operator complexity beyond scrambling},
  author={Barb{\'o}n, JLF and Rabinovici, Eliezer and Shir, R and Sinha, R},
  journal={Journal of High Energy Physics},
  volume={2019},
  number={10},
  pages={1--25},
  year={2019},
  publisher={Springer},
  doi = {https://doi.org/10.1007/JHEP10(2019)264}
}

@Article{Bhattacharjee2022saddle,
author={Bhattacharjee, Budhaditya
and Cao, Xiangyu
and Nandy, Pratik
and Pathak, Tanay},
title={Krylov complexity in saddle-dominated scrambling},
journal={Journal of High Energy Physics},
year={2022},
month={May},
day={25},
volume={2022},
number={5},
pages={174},
issn={1029-8479},
doi={10.1007/JHEP05(2022)174},
url={https://doi.org/10.1007/JHEP05(2022)174}
}

@Article{Bhattacharya2022arnoldi,
author={Bhattacharya, Aranya
and Nandy, Pratik
and Nath, Pingal Pratyush
and Sahu, Himanshu},
title={Operator growth and {K}rylov construction in dissipative open quantum systems},
journal={Journal of High Energy Physics},
year={2022},
month={Dec},
day={14},
volume={2022},
number={12},
pages={81},
issn={1029-8479},
doi={10.1007/JHEP12(2022)081},
url={https://doi.org/10.1007/JHEP12(2022)081}
}

@Article{Bhattacharjee2023dissSYK,
author={Bhattacharjee, Budhaditya
and Cao, Xiangyu
and Nandy, Pratik
and Pathak, Tanay},
title={Operator growth in open quantum systems: lessons from the dissipative {SYK}},
journal={Journal of High Energy Physics},
year={2023},
month={Mar},
day={08},
volume={2023},
number={3},
pages={54},
issn={1029-8479},
doi={10.1007/JHEP03(2023)054},
url={https://doi.org/10.1007/JHEP03(2023)054}
}

@article{Nizami2023Floquet,
  title = {Krylov construction and complexity for driven quantum systems},
  author = {Nizami, Amin A. and Shrestha, Ankit W.},
  journal = {Phys. Rev. E},
  volume = {108},
  issue = {5},
  pages = {054222},
  numpages = {11},
  year = {2023},
  month = {Nov},
  publisher = {American Physical Society},
  doi = {10.1103/PhysRevE.108.054222},
  url = {https://link.aps.org/doi/10.1103/PhysRevE.108.054222}
}

@article{Nizami2024Floquet,
  title = {Spread complexity and quantum chaos for periodically driven spin chains},
  author = {Nizami, Amin A. and Shrestha, Ankit W.},
  journal = {Phys. Rev. E},
  volume = {110},
  issue = {3},
  pages = {034201},
  numpages = {13},
  year = {2024},
  month = {Sep},
  publisher = {American Physical Society},
  doi = {10.1103/PhysRevE.110.034201},
  url = {https://link.aps.org/doi/10.1103/PhysRevE.110.034201}
}

@Article{Bhattacharya2023biLanczos,
author={Bhattacharya, Aranya
and Nandy, Pratik
and Nath, Pingal Pratyush
and Sahu, Himanshu},
title={On {K}rylov complexity in open systems: an approach via bi-{L}anczos algorithm},
journal={Journal of High Energy Physics},
year={2023},
month={Dec},
day={12},
volume={2023},
number={12},
pages={66},
issn={1029-8479},
doi={10.1007/JHEP12(2023)066},
url={https://doi.org/10.1007/JHEP12(2023)066}
}

@article{Zhou2025manybodychaos,
  title = {Diagnosing quantum many-body chaos in non-{H}ermitian quantum spin chain via {K}rylov complexity},
  author = {Zhou, Yijia and Xia, Wei and Li, Lin and Li, Weibin},
  journal = {Phys. Rev. Res.},
  volume = {7},
  issue = {3},
  pages = {033281},
  numpages = {12},
  year = {2025},
  month = {Sep},
  publisher = {American Physical Society},
  doi = {10.1103/fw62-j2n9},
  url = {https://link.aps.org/doi/10.1103/fw62-j2n9}
}

@article{baggioli2026nonhermitian,
      title={Quantum Chaos Diagnostics for non-{H}ermitian Systems from Bi-{L}anczos {K}rylov Dynamics}, 
      author={Matteo Baggioli and Kyoung-Bum Huh and Hyun-Sik Jeong and Xuhao Jiang and Keun-Young Kim and Juan F. Pedraza},
      year={2026},
      eprint={2508.13956},
      archivePrefix={arXiv},
      url={https://arxiv.org/abs/2508.13956}, 
      journal = {}
}

@article{bhattacharyya2025kcopenquantum,
      title={Krylov Complexity for Open Quantum System: Dissipation and Decoherence}, 
      author={Arpan Bhattacharyya and Sayed Gool and S. Shajidul Haque},
      year={2025},
      eprint={2509.14810},
      archivePrefix={arXiv},
      url={https://arxiv.org/abs/2509.14810}, 
      journal = {}
}

@article{Ibanez2011ChirpedPulses,
  title = {Interaction of strongly chirped pulses with two-level atoms},
  author = {Ib\'a\~nez, S. and Peralta Conde, A. and Gu\'ery-Odelin, D. and Muga, J. G.},
  journal = {Phys. Rev. A},
  volume = {84},
  issue = {1},
  pages = {013428},
  numpages = {6},
  year = {2011},
  month = {Jul},
  publisher = {American Physical Society},
  doi = {10.1103/PhysRevA.84.013428},
  url = {https://link.aps.org/doi/10.1103/PhysRevA.84.013428}
}

@article{HatanoNelson1996,
  title = {Localization Transitions in Non-{H}ermitian Quantum Mechanics},
  author = {Hatano, Naomichi and Nelson, David R.},
  journal = {Phys. Rev. Lett.},
  volume = {77},
  issue = {3},
  pages = {570--573},
  numpages = {0},
  year = {1996},
  month = {Jul},
  publisher = {American Physical Society},
  doi = {10.1103/PhysRevLett.77.570},
  url = {https://link.aps.org/doi/10.1103/PhysRevLett.77.570}
}

@article{HatanoNelson1997,
  title = {Vortex pinning and non-{H}ermitian quantum mechanics},
  author = {Hatano, Naomichi and Nelson, David R.},
  journal = {Phys. Rev. B},
  volume = {56},
  issue = {14},
  pages = {8651--8673},
  numpages = {0},
  year = {1997},
  month = {Oct},
  publisher = {American Physical Society},
  doi = {10.1103/PhysRevB.56.8651},
  url = {https://link.aps.org/doi/10.1103/PhysRevB.56.8651}
}

@article{Grabarits2026gaps,
  title = {Fighting Exponentially Small Gaps by Counterdiabatic Driving},
  author = {Grabarits, Andr\'as and Balducci, Federico and del Campo, Adolfo},
  journal = {PRX Quantum},
  volume = {7},
  issue = {1},
  pages = {010322},
  numpages = {43},
  year = {2026},
  month = {Feb},
  publisher = {American Physical Society},
  doi = {10.1103/tgzt-dy3h},
  url = {https://link.aps.org/doi/10.1103/tgzt-dy3h}
}

@article{DoraQuench2023,
  title = {Quantum quench dynamics in the Luttinger liquid phase of the {H}atano-{N}elson model},
  author = {D\'ora, Bal\'azs and Werner, Mikl\'os Antal and Moca, Catalin Pascu},
  journal = {Phys. Rev. B},
  volume = {108},
  issue = {3},
  pages = {035104},
  numpages = {8},
  year = {2023},
  month = {Jul},
  publisher = {American Physical Society},
  doi = {10.1103/PhysRevB.108.035104},
  url = {https://link.aps.org/doi/10.1103/PhysRevB.108.035104}
}

@article{Bender_PT,
  title = {Real Spectra in Non-{H}ermitian {H}amiltonians Having {PT} {S}ymmetry},
  author = {Bender, Carl M. and Boettcher, Stefan},
  journal = {Phys. Rev. Lett.},
  volume = {80},
  issue = {24},
  pages = {5243--5246},
  numpages = {0},
  year = {1998},
  month = {Jun},
  publisher = {American Physical Society},
  doi = {10.1103/PhysRevLett.80.5243},
  url = {https://link.aps.org/doi/10.1103/PhysRevLett.80.5243}}

@article{Kattel_PT_2023,
doi = {10.1088/1751-8121/ace56e},
url = {https://doi.org/10.1088/1751-8121/ace56e},
year = {2023},
month = {jul},
publisher = {IOP Publishing},
volume = {56},
number = {32},
pages = {325001},
author = {Kattel, Pradip and Pasnoori, Parameshwar R and Andrei, Natan},
title = {Exact solution of a non-{H}ermitian {PT}-symmetric spin chain},
journal = {Journal of Physics A: Mathematical and Theoretical}
}

@article{Ashida_NHPhysics,
author = {Yuto Ashida and Zongping Gong and Masahito Ueda},
title = {Non-{H}ermitian physics},
journal = {Advances in Physics},
volume = {69},
number = {3},
pages = {249--435},
year = {2020},
publisher = {Taylor \& Francis},
doi = {10.1080/00018732.2021.1876991},
URL = {https://doi.org/10.1080/00018732.2021.1876991}
}

@misc{code,
author = {Shrestha, Ankit W. and Bhattacharjee, Budhaditya and del Campo, Adolfo},
year = {2026},
howpublished = {\url{https://github.com/A-Wenju/STA_non_Hermitian_Krylov}},
    url = {https://github.com/A-Wenju/STA_non_Hermitian_Krylov/tree/main}
}

@article{Alipour2020shortcutsto,
  doi = {10.22331/q-2020-09-28-336},
  url = {https://doi.org/10.22331/q-2020-09-28-336},
  title = {Shortcuts to {A}diabaticity in {D}riven {O}pen {Q}uantum {S}ystems: {B}alanced {G}ain and {L}oss and {N}on-{M}arkovian {E}volution},
  author = {Alipour, Sahar and Chenu, Aurelia and Rezakhani, Ali T. and del Campo, Adolfo},
  journal = {{Quantum}},
  issn = {2521-327X},
  publisher = {{Verein zur F{\"{o}}rderung des Open Access Publizierens in den Quantenwissenschaften}},
  volume = {4},
  pages = {336},
  month = sep,
  year = {2020}
}

@article{Hatomura2021Controlling,
  title = {Controlling and exploring quantum systems by algebraic expression of adiabatic gauge potential},
  author = {Hatomura, Takuya and Takahashi, Kazutaka},
  journal = {Phys. Rev. A},
  volume = {103},
  issue = {1},
  pages = {012220},
  numpages = {8},
  year = {2021},
  month = {Jan},
  publisher = {American Physical Society},
  doi = {10.1103/PhysRevA.103.012220},
  url = {https://link.aps.org/doi/10.1103/PhysRevA.103.012220}
}

@article{Yang2022hidden,
    doi = {10.1088/1367-2630/ac652f},
    url = {https://doi.org/10.1088/1367-2630/ac652f},
    year = {2022},
    month = {apr},
    publisher = {IOP Publishing},
    volume = {24},
    number = {4},
    pages = {043046},
    author = {Yang, Fei and Wang, Heng and Yang, Meng-Lei and Guo, Cui-Xian and Wang, Xiao-Ran and Sun, Gao-Yong and Kou, Su-Peng},
    title = {Hidden continuous quantum phase transition without gap closing in non-{H}ermitian transverse Ising model},
    journal = {New Journal of Physics}
}

@article{Lu2024msnybody,
  title = {Many-body phase transitions in a non-{H}ermitian Ising chain},
  author = {Lu, Chao-Ze and Deng, Xiaolong and Kou, Su-Peng and Sun, Gaoyong},
  journal = {Phys. Rev. B},
  volume = {110},
  issue = {1},
  pages = {014441},
  numpages = {7},
  year = {2024},
  month = {Jul},
  publisher = {American Physical Society},
  doi = {10.1103/PhysRevB.110.014441},
  url = {https://link.aps.org/doi/10.1103/PhysRevB.110.014441}
}

@article{Sun2021biorthogonal,
    author={Sun, Gaoyong
    and Tang, Jia-Chen
    and Kou, Su-Peng},
    title={Biorthogonal quantum criticality in non-{H}ermitian many-body systems},
    journal={Frontiers of Physics},
    year={2021},
    month={Nov},
    day={19},
    volume={17},
    number={3},
    pages={33502},
    issn={2095-0470},
    doi={10.1007/s11467-021-1126-1},
    url={https://doi.org/10.1007/s11467-021-1126-1}
}

@article{usmani1994inversion,
    title = {Inversion of a tridiagonal jacobi matrix},
    journal = {Linear Algebra and its Applications},
    volume = {212-213},
    pages = {413-414},
    year = {1994},
    issn = {0024-3795},
    doi = {https://doi.org/10.1016/0024-3795(94)90414-6},
    url = {https://www.sciencedirect.com/science/article/pii/0024379594904146},
    author = {Riaz A. Usmani}
}

\end{document}